\pdfoutput=1

\documentclass[twocolumn]{aastex631}
\usepackage{epstopdf}
\usepackage{xspace}
\usepackage{graphicx}
\usepackage{amssymb}
\usepackage{amsmath}
\usepackage{natbib}
\usepackage{float}
\usepackage{hyperref}

\shorttitle{X-ray Emission of NGC\,3894}
\shortauthors{Balasubramaniam et al.}

\begin{document}

\title{X-ray Emission of the $\gamma$-ray--Loud Young Radio Galaxy NGC\,3894}

\correspondingauthor{K.~Balasubramaniam}
\email{karthik.balasubramaniam@doctoral.uj.edu.pl}

\author{K.~Balasubramaniam}
\affiliation{Astronomical Observatory of the Jagiellonian University, ul. Orla 171, 30-244 Krak\'ow, Poland}

\author{\L .~Stawarz}
\affiliation{Astronomical Observatory of the Jagiellonian University, ul. Orla 171, 30-244 Krak\'ow, Poland}

\author{C.~C.~Cheung}
\affiliation{Naval Research Laboratory, Space Science Division, Washington, DC 20375, USA}

\author{M.~Sobolewska},
\affiliation{Harvard Smithsonian Center for Astrophysics, 60 Garden Street, Cambridge, MA 02138, USA}

\author{V.~Marchenko}
\affiliation{Astronomical Observatory of the Jagiellonian University, ul. Orla 171, 30-244 Krak\'ow, Poland}

\author{R.~Thimmappa}
\affiliation{Astronomical Observatory of the Jagiellonian University, ul. Orla 171, 30-244 Krak\'ow, Poland}

\author{D.~{\L}.~Kr\'ol}
\affiliation{Astronomical Observatory of the Jagiellonian University, ul. Orla 171, 30-244 Krak\'ow, Poland}

\author{G.~Migliori}
\affiliation{Department of Physics and Astronomy, University of Bologna, Via Gobetti 93/2, I-40129 Bologna, Italy; INAF-Institute of Radio Astronomy, Bologna, Via Gobetti 101, I-40129 Bologna, Italy}

\author{A.~Siemiginowska},
\affiliation{Harvard Smithsonian Center for Astrophysics, 60 Garden Street, Cambridge, MA 02138, USA}

\begin{abstract}
The radio source 1146+596 is hosted by an elliptical/S0 galaxy NGC\,3894, with a low-luminosity active nucleus. The radio structure is compact, suggesting a very young age of the jets in the system. Recently, the source has been confirmed as a high-energy (HE, $>0.1$\,GeV) $\gamma$-ray emitter, in the most recent accumulation of the {\it Fermi} Large Area Telescope (LAT) data. Here we report on the analysis of the archival {\it Chandra} X-ray Observatory data for the central part of the galaxy, consisting of a single 40\,ksec-long exposure. We have found that the core spectrum is best fitted by a combination of an ionized thermal plasma with the temperature of $\simeq 0.8$\,keV, and a moderately absorbed power-law component (photon index $\Gamma = 1.4\pm 0.4$, hydrogen column density $N_{\rm H}/10^{22}$\,cm$^{-2}$\,$= 2.4\pm 0.7$). We have also detected the iron K$\alpha$ line at $6.5\pm 0.1$\,keV, with a large equivalent width of EW\,$= 1.0_{-0.5}^{+0.9}$\,keV. Based on the simulations of the {\it Chandra}'s Point Spread Function (PSF), we have concluded that, while the soft thermal component is extended on the scale of the galaxy host, the hard X-ray emission within the narrow photon energy range 6.0--7.0\,keV originates within the unresolved core (effectively the central kpc radius). The line is therefore indicative of the X-ray reflection from a cold neutral gas in the central regions of NGC\,3894. We discuss the implications of our findings in the context of the X-ray Baldwin effect. NGC\,3894 is the first young radio galaxy detected in HE $\gamma$-rays with the iron K$\alpha$ line.
\end{abstract}


\section{Introduction}
\label{sec:intro}

The elliptical/lenicular galaxy NGC\,3894, a member of a non-interacting pair located at RA(J2000)=11h48m50.36s and Dec(J2000)=+59$^{\circ}$24$^{\prime}$56.43$^{\prime\prime}$ \citep{nilson73,devaucoulers91}, hosts a compact radio source 1146+596 \citep{Condon78}. The first Very-Long-Baseline Interferometry (VLBI) of the system at 5\,GHz \citep{wrobel85}, suggested an asymmetric core-jet morphology on parsec scales. Subsequent VLBI monitoring of the asymmetric double-lobed structure revealed properties more consistent with a `Compact Symmetric Object' (CSO) classification \citep[see][for reviews]{odea98,odea21}. In particular, the 1981--1996 VLBI monitoring data analyzed by \citet{taylor98}, indicated that the twin jets in the system were characterized by only sub-relativistic expansion, and oriented at rather large, $\sim 50^{\circ}$ angle from the line-of-sight. As such, 1146+596/NGC\,3894 became recognized as one of the nearest and the youngest low-power radio galaxy.

\begin{figure*}[!th]
\centering
\includegraphics[width=0.75\textwidth]{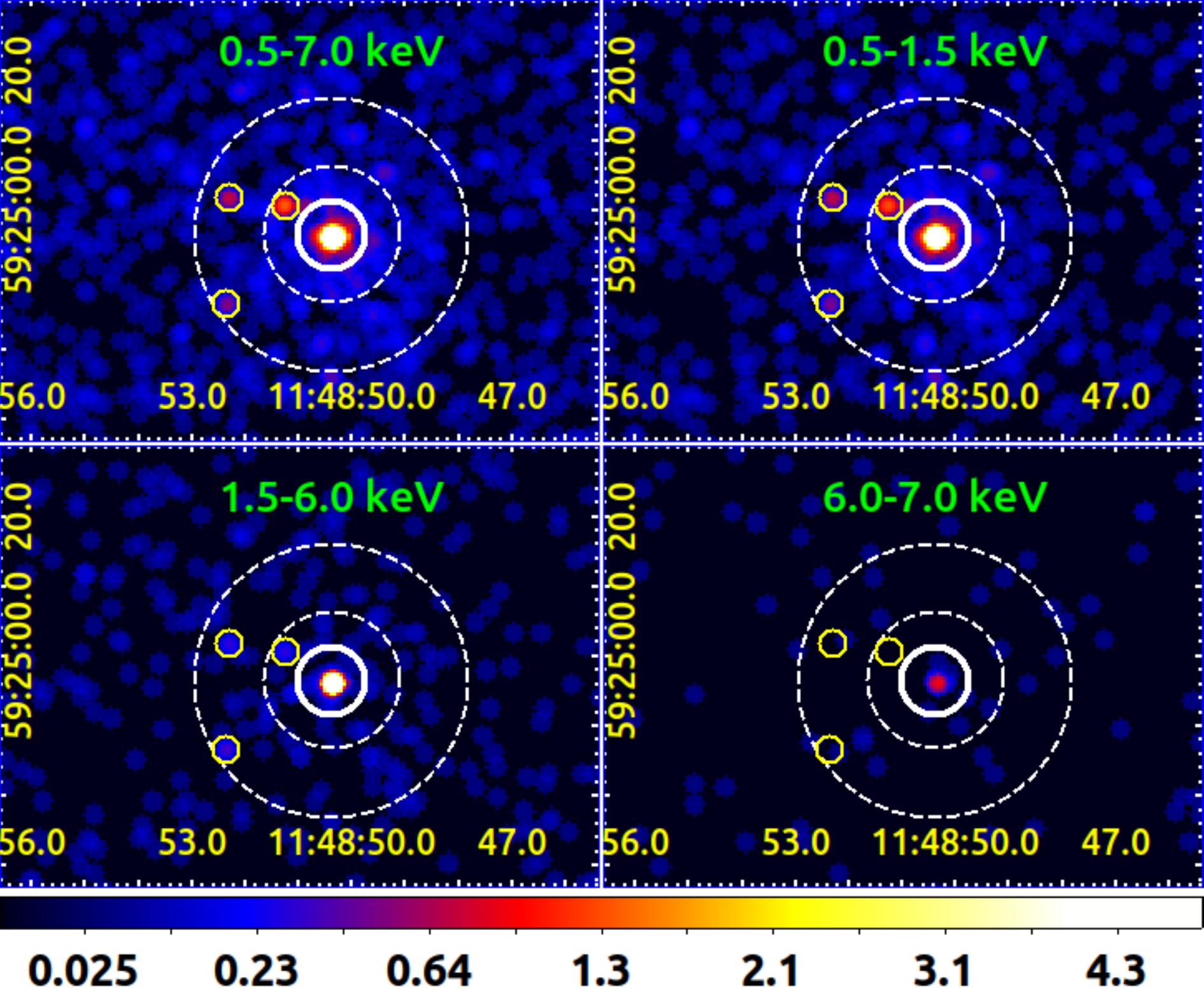}
\caption{\textit{Chandra} images of NGC\,3894, smoothed with $3\sigma$ kernel, within the full \textit{Chandra} photon energy range 0.5--7.0\,keV (upper left panel), the soft band 0.5--1.5\,keV (upper right), the medium band 1.5--6.0\,keV (bottom left), and the hard band 6.0--7.0\,keV (bottom right). The applied core extraction region with the 10\,px\,$\simeq 5^{\prime\prime}$ radius is denoted in all the panels by a white solid circle; the background for the core spectral analysis was chosen as an annulus with the inner and outer radii of $10^{\prime\prime}$ and $20^{\prime\prime}$, respectively, as represented in the panels by white dashed circles (with the two prominent point sources, marked in the figure by yellow solid circles, removed).} 
\label{fig:chandra}
\end{figure*}

Very recently, \citet{Principe2020}, who analyzed the earlier VLBI data together with five additional Very Long Baseline Array (VLBA) datasets, confirmed the youth scenario for 1146+596, with the projected size of $\sim 4$\,pc and the jet advance speed $\sim (0.3-0.4)\,c$, but at the same time a considerably smaller jet viewing angle of $\sim 10^{\circ}-20^{\circ}$. The corresponding dynamical age of the radio structure reads as $\sim 60$\,years.

Redshifted HI absorption has been detected towards both the approaching and receding jets of 1146+596 \citep{vanGorkom89,peck98,gupta06}. Distinct absorption features have been interpreted as related to either gaseous clouds falling onto the nucleus, or a circumnuclear hot dusty torus. The corresponding neutral hydrogen column densities has been estimated as $N_{\rm HI} \simeq 4 \times 10^{20}$\,cm$^{-2}$ on average, with the $(2-14)\times 10^{20}$\,cm$^{-2}$ range found for distinct components of the radio structure \citep[see also][]{perlman01,emonts10}.

The infrared (IR) studies of NGC\,3894 using various telescopes, such as the Infrared Astronomical Satellite (IRAS), the {\it Spitzer} Space Telescope, and the Wide-field Infrared Survey Explorer \citep[WISE;][]{wil10,Kosmaczewski2020}, reinforced the `Low-ionization nuclear emission-line region' (LINER) classification of the nucleus \citep[see in this context][]{Condon88,kim89,Goncalves04}, and indicated a rather low star formation rate in the host, at the level of $\sim 0.5 M_{\odot}$\,yr$^{-1}$ .

At optical frequencies, the high-resolution imaging performed with the telescopes of Lick Observatory and also the {\it Hubble} Space Telescope \citep{kim89,perlman01}, revealed the presence of dust lanes and ionized gas distributed along the major axis of the host, nearly perpendicular to the radio axes, likely related to the circumnuclear hot dusty torus.

\begin{figure*}[!th]
\centering
\includegraphics[width=0.63\textwidth]{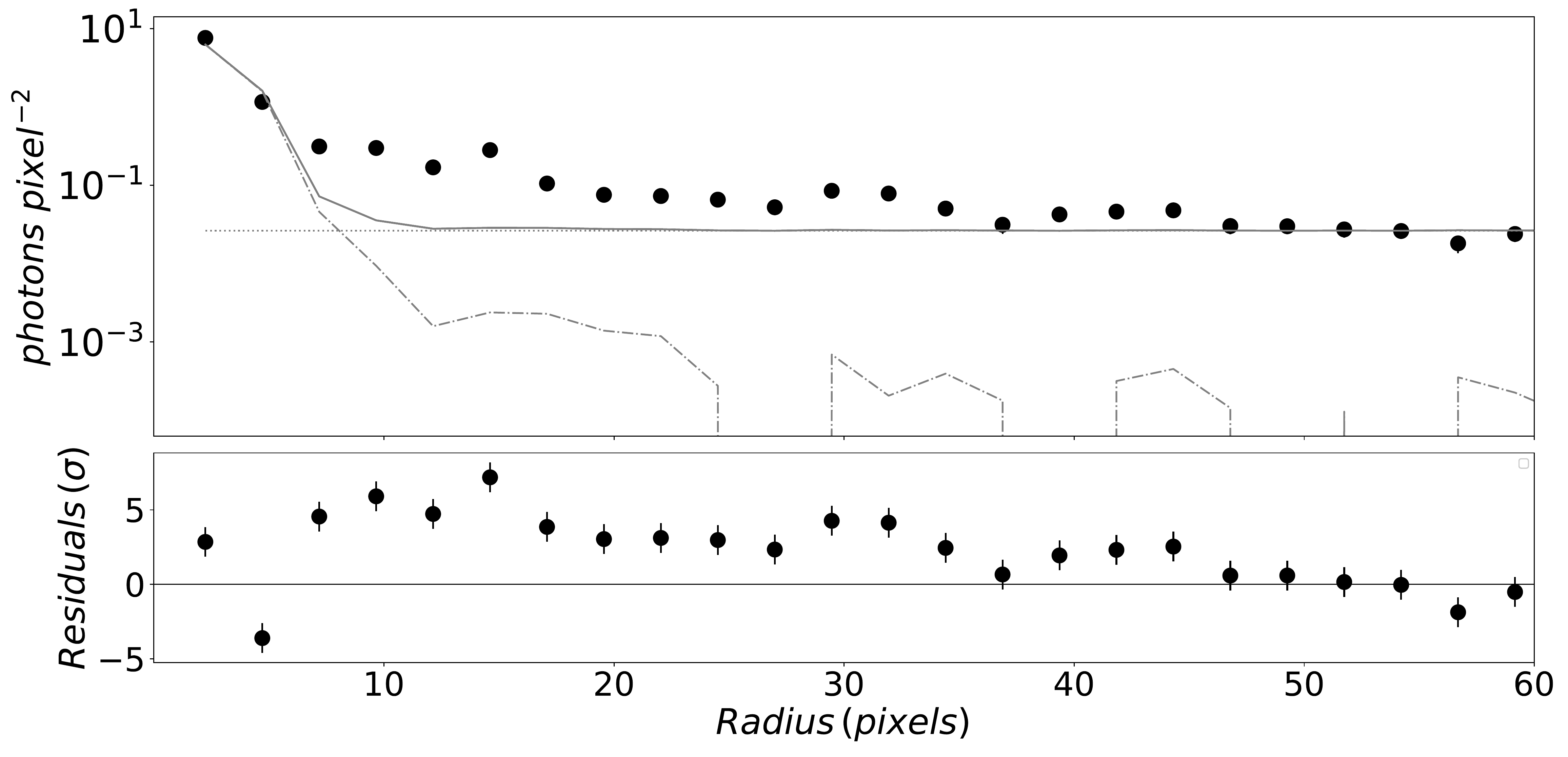}
\includegraphics[width=0.63\textwidth]{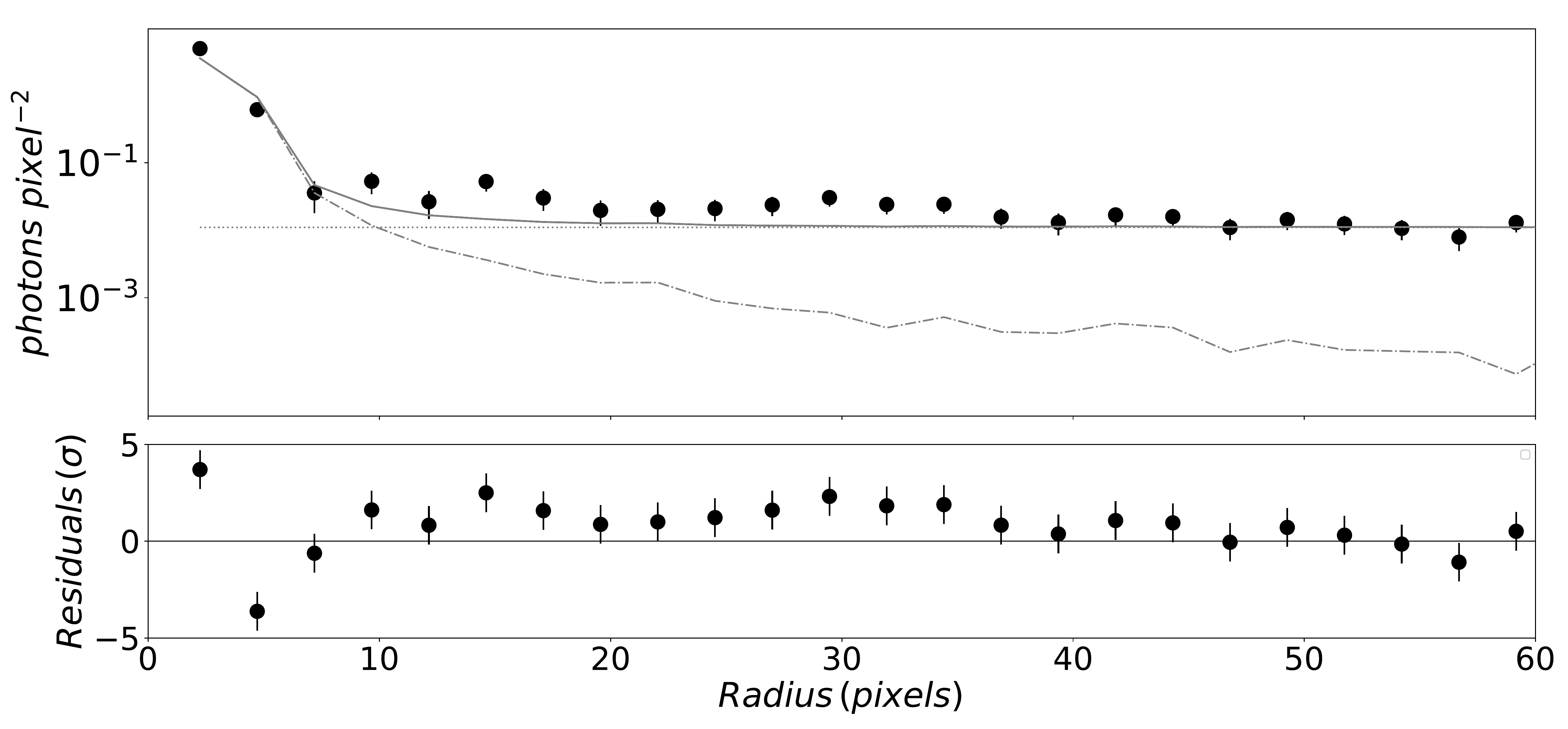}
\includegraphics[width=0.63\textwidth]{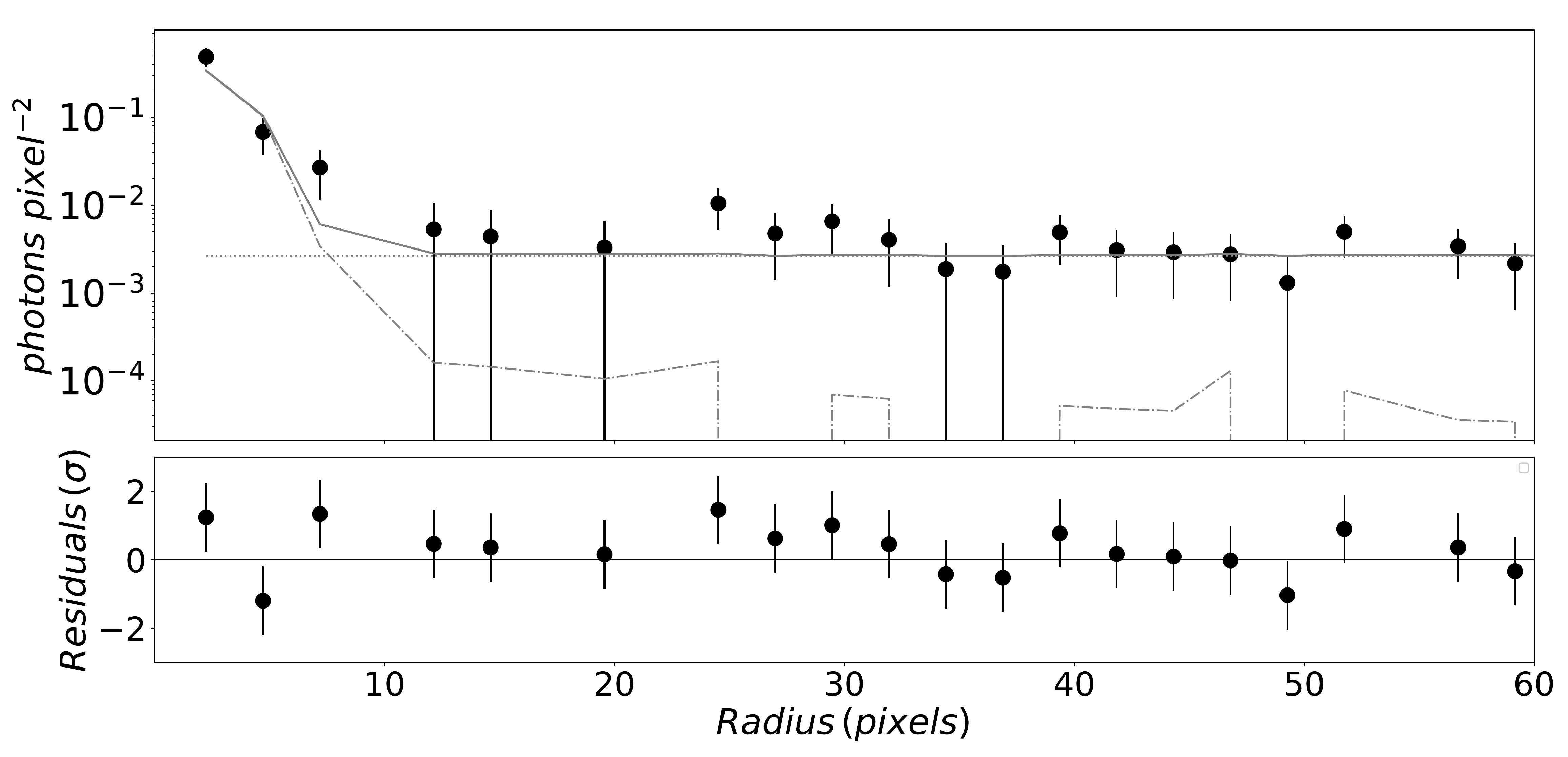}
 \caption{The X-ray surface brightness profiles of NGC\,3894, within the narrow photon energy ranges 0.5--1.5\,keV (soft band; upper panel), 1.5--6.0\,keV (medium band; middle panel), and 6.0--7.0\,keV (hard band; lower panel). Long dashed grey curves denote the table models for the core PSFs simulated specifically for the corresponding bands, while the horizontal dashed lines denote the constant background levels. Residuals of the surface brightness profile fitting with the PSF+background models are given in each panel.}
 \label{fig:SB}
\end{figure*} 

Finally, NGC\,3894 was recently identified as a counterpart to the HE $\gamma$-ray source 4FGL\,J1149.0+5924 listed in the
{\it Fermi}-LAT 8-year Source Catalog \citep[4FGL;][]{4FGL,Principe2020}. As such, it is one of only several young radio galaxies detected in the HE $\gamma$-ray range \citep[see][]{McConville11,Muller14,Migliori14,Migliori16,DAmmando16,Lister20}.

All in all, one can conclude that  1146+596/NGC\,3894 is a relevant source, well studied and characterized at radio, IR, optical, and HE $\gamma$-ray ranges, but at the same time lacking any detailed X-ray spectroscopy and imaging. Here we therefore analyzed the archival {\it Chandra} data for the system, consisting of a single 40\,ksec exposure with the Advanced CCD Imaging Spectrometer (ACIS) detector. Along with presenting the results of the {\it Chandra} data analysis, we also update the broad-band Spectral Energy Distribution (SED) of the source.

\begin{figure}[!th]
\centering
\includegraphics[width=\columnwidth]{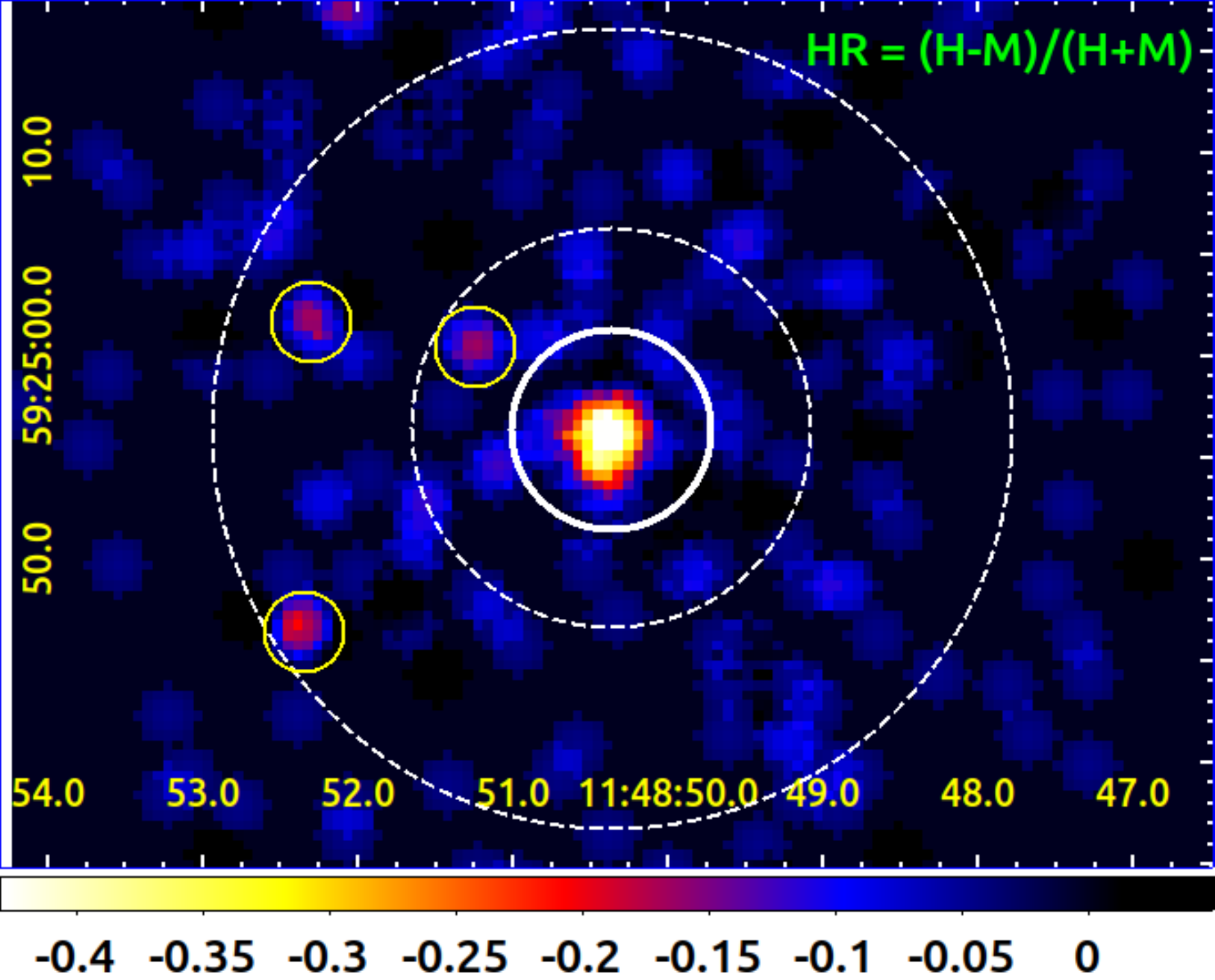}
\caption{The hardness ratio ACIS-S image of the central parts of NGC\,3894, defined as HR=(H-M)/(H+M) for the medium 1.5--6.0\,keV (M) and hard 6.0--7.0\,keV (H) bands.} 
\label{fig:HR}
\end{figure}

Throughout the paper we assume modern $\Lambda$CDM cosmology with $H_{0}=70$\,km\,s$^{-1}$\,Mpc$^{-1}$, $\Omega _{\rm m}=0.3$, and $\Omega _{\Lambda}=0.7$, so that the luminosity distance of the source for the given redshift $z = 0.01075$ is $d_{\rm L} = 50.1$\,Mpc, and the conversion scale reads as 0.238\,kpc/arcsec.

\section{Chandra Data Analysis} 
\label{sec:chandra}

The {\it Chandra} ACIS-S observation of the galaxy NGC\,3894 were conducted in 2009 July, during the Observing Cycle 10 (PI: E. Perlman), with the exposure totalling to 38.4\,ksec. The source was placed at the aim-point on the back-illuminated ACIS CCD. The data were collected in the VFAINT mode and the observation was made in the timed exposure mode. The target was detected with the total number of net counts of 457 within the 10\,px\,$\simeq 5^{\prime\prime}$ radius from the core, in the full photon energy range 0.5--7.0\,keV utilized in our analysis. Figure\,\ref{fig:chandra} presents the resulting \textit{Chandra} images of NGC\,3894, smoothed with $3\sigma$ kernel, within the full 0.5--7.0\,keV range, the soft band 0.5--1.5\,keV, the medium band 1.5--6.0\,keV, and the hard band 6.0--7.0\,keV.

The X-ray data analysis was performed with the CIAO v.4.13 software \citep{Fruscione06} using CALDB v.4.8.5. We processed the data by running the CIAO tool \texttt{chandra$\_$repro}. Spectral modeling was done in \texttt{XSPEC}\footnote{\url{https://heasarc.gsfc.nasa.gov/xanadu/xspec}} version 12.11.1 \citep{Arnaud1996}, using the Chi-square fitting statistics and the Nelder-Mead optimization method, and also with \texttt{Sherpa} \citep{Freeman01}, when working on unbinned data, using the Cash statistics and the Nelder-Mead optimization method. The uncertainties of spectral paramateres provided in the text, are $1\sigma$ errors. Below in \S\,\ref{sec:profile} we present the detailed PSF simulations and the surface brightness profile analysis in the narrow energy ranges (soft, medium, and hard bands), due to the fact that the source spectrum appears relatively hard, and also due to the presence of a strong iron fluorescence; next in \S\,\ref{sec:spectrum} we present spectral modelling.

\begin{deluxetable*}{cccccccccc}[!th] 
\tabletypesize{\scriptsize}
\tablecaption{Spectral modelling of the binned nuclear spectrum of NGC\,3894}
\label{tab:spec}
\tablewidth{0pt}
\tablehead{
\colhead{Model} & \colhead{$N_{\rm H}$}  & \colhead{$kT$} & \colhead{APEC norm.$^{\dagger}$} & \colhead{$E_{\ell}$} & \colhead{$\sigma_{\ell}$} & \colhead{Line norm.$^{\dagger}$} & \colhead{$\Gamma$} & \colhead{PL norm.$^{\dagger}$}  & \colhead{$\chi^{2}$/DOF} \\
\colhead{~~} & \colhead{$\times 10^{22}$\,cm$^{-2}$} & \colhead{keV} & \colhead{$\times 10^{-6}$\,\texttt{XSPEC}} & \colhead{keV} &  \colhead{keV} & \colhead{$\times 10^{-6}$\,\texttt{XSPEC}} & \colhead{~~} & \colhead{$\times 10^{-6}$\,\texttt{XSPEC}} & \colhead{~~} \\
\colhead{(1)} & \colhead{(2)} & \colhead{(3)} & \colhead{(4)} & \colhead{(5)} & \colhead{(6)} & \colhead{(7)} & \colhead{(8)} & \colhead{(9)} & \colhead{(10)} 
}
\startdata
(i) & $2.0 \pm 0.6$ & $0.8 \pm 0.1$ & $3.4 \pm 0.4$ & --- & --- & --- & $1.1 \pm 0.4$ & $20 \pm 11$ & $64.27/72$\\
(ii) & $2.2 \pm 0.7$ & $0.8 \pm 0.1$ & $3.5 \pm 0.4$ & $6.5 \pm 0.1$ & $0.01$ (frozen) & $1.3 \pm  0.6$ & $1.3 \pm 0.4$ & $24 \pm 14$ & $59.21/70$\\
(iii) & $2.4 \pm 0.7$ & $0.8 \pm 0.1$ & $3.5 \pm 0.4$ & $6.5 \pm 0.1$ & $0.12 \pm 0.08$ & $2.0 \pm  0.9$ & $1.4 \pm 0.4$ & $29 \pm 17$ & $57.90/69$\\
\enddata
\tablecomments{Col(1) --- model fit; Col(2) --- intrinsic hydrogen column density; Col(3--4) --- temperature and normalization of the collisionally ionized thermal plasma; Col(5--7) --- Gaussian line energy, width, and normalization; Col(8--9) --- photon index and normalization of the power-law emission component; Col(10) --- $\chi^2$ fitting statistics/DOF. All the models include Galactic hydrogen column density $N_{\rm H,\,Gal} = 1.83 \times 10^{20}$\,cm$^{-2}$; thermal component assumes solar abundance. The 0.5--7.0\,keV unabsorbed flux of the power-law component reads as $F_{\rm 0.5-7.0\,keV} \simeq 2\times 10^{-13}$\,erg\,cm$^{-2}$\,s$^{-1}$. The 6.5\,keV line equivalent width ($1\sigma$ range) reads as EW\,$= 0.6 \, (0.4-1.1)$\,keV for the fit with frozen $\sigma_{\ell}=0.01$\,keV, and EW\,$= 1.0\,(0.5-1.9)$\,keV for $\sigma_{\ell}$ set free. $^{\dagger}$ The \texttt{XSPEC} normalizations are: (a) for the APEC component $10^{-14} (1+z)^2 n_e n_{\rm H} V/4\pi d_{\rm L}^2$, assuming uniform ionized plasma with electron and H number densities $n_e$ and $n_{\rm H}$, respectively, and volume $V$, all in cgs units; (b) for the Gaussian line component total photons/cm$^2$/s in the line; (c) for the power-law component photons/keV/cm$^2$/s at 1\,keV.}
\end{deluxetable*}

\subsection{Surface Brightness Profiles}
\label{sec:profile}

In order to correct for the {\it Chandra} dithering motion, the limited size of detector pixels, and detector effects, we performed the High Resolution Mirror Assembly (HRMA) PSF simulations for the NGC\,3894 nucleus, using the Chandra Ray Tracer (ChaRT) online tool \citep{Carter03}\footnote{\url{http://cxc.harvard.edu/ciao/PSFs/chart2/runchart.html}}. For these, the source spectrum file was uploaded to ChaRT, obtaining a set of rays, which were next projected onto the detector plane with the MARX software \citep{Davis12}\footnote{\url{https://space.mit.edu/cxc/marx}}. The resulting PSF files were normalized to the observed count rate, and filtered with the source region at bin factor 1. 

Because of limited photon statistics we are dealing with here, one should expect possibly quite substantial differences in each particular realisation of the PSF, due to random photon fluctuations. For this reason, we have repeated the PSF simulations 50 times, and then averaged them, obtaining in this way the final PSF model that was used for the surface brightness profile fitting. The PSF simulations as described above have been performed separately for the three bands, namely the soft band 0.5--1.5\,keV, the medium band 1.5--6.0\,keV, and the hard band 6.0--7.0\,keV. We note an extremely low photon statistics in the lattermost case, amounting to only $26$ net photons within the $<10$\,px\,$\simeq 5^{\prime\prime}$ source extraction region. Despite this obstacle, the 6.0--7.0\,keV PSF could still be characterized reasonably well.

After the extensive PSF simulations, we extracted observed counts from the exposure-corrected Chandra map in a concentric stack of annular regions centred on the galactic nucleus up to $\sim 60$\,px (without removing any point sources), and fit the resulting radial profile of the X-ray surface brightness of the target with the model including the core PSF, and a constant background. During the fitting procedure, the PSF (table model) normalization, as well as the background amplitudes, were in all the cases set free. The corresponding best-fit lines along with the residuals, are given in Figure\,\ref{fig:SB}.

As shown in the figure, in the case of the soft {\it Chandra} band, a clear source extension beyond the core PSF is seen, manifesting as positive residuals in the fit at the level of (and above) $3\sigma$, up to $\sim 30$\,px\,$\simeq 3.5$\,kpc from the core. This extension cannot be ascribed solely to the three identified point sources, although their contribution to the surface brightness can be seen in the radial profile of the soft-band emission around 15\,px and 30--35\,px from the core. In the medium band, on the other hand, no significant extension beyond the central PSF can be found, except of a minor excess around the position of the three point sources (positive residual $<3\sigma$). In the hard band, finally, no source extension --- either due to a diffuse component, or due to the identified point sources --- is present beyond the central PSF (effectively the central $<5^{\prime\prime}\simeq 1.2$\,kpc radius), although a very limited photon statistics in this band should be kept in mind.

\begin{figure*}[!th]
\centering
\includegraphics[width=0.75\textwidth]{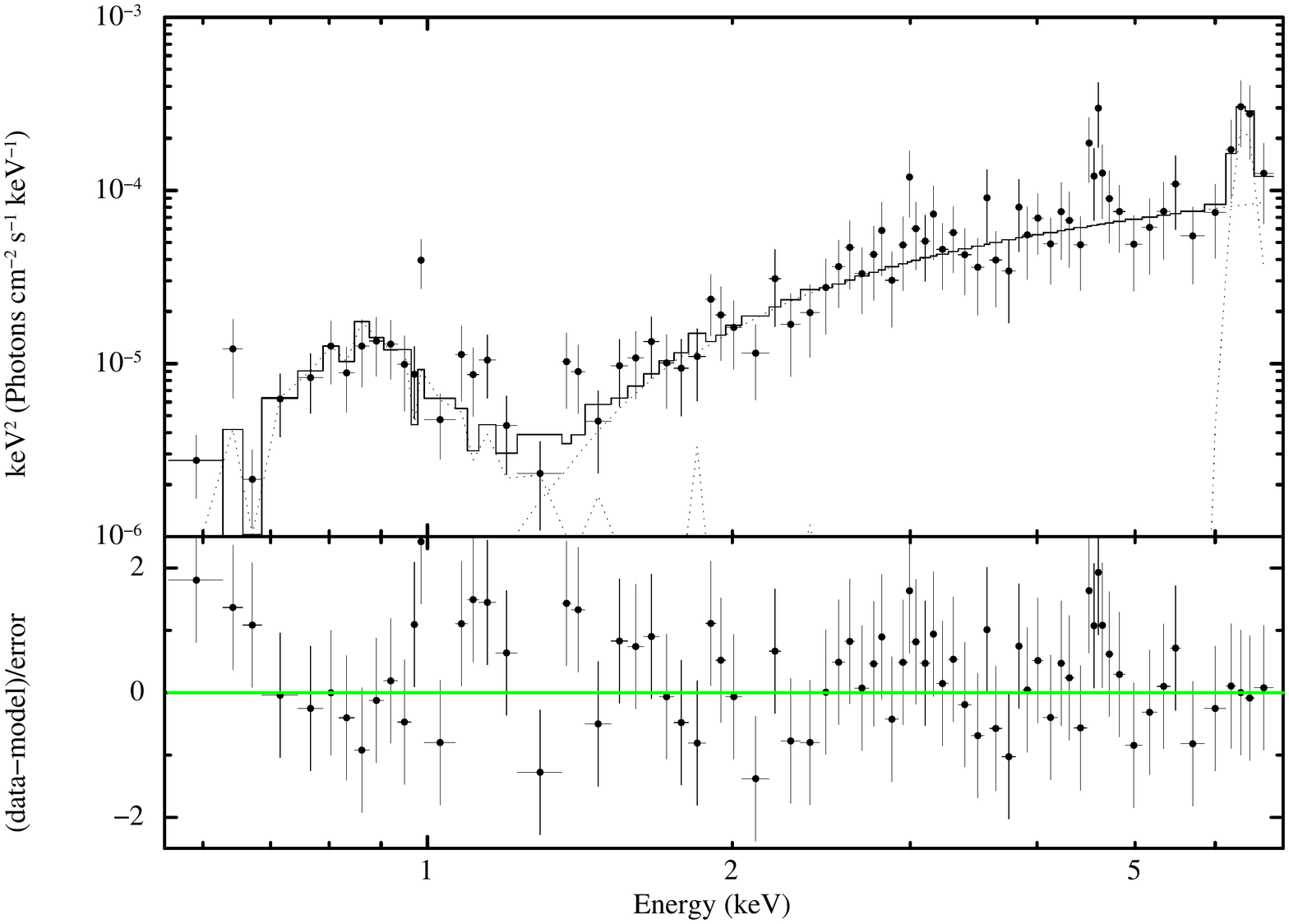}
\caption{Background-subtracted $0.5-7$\,keV {\it Chandra} spectrum of the unresolved X-ray core in NGC\,3894, grouped with minimum 5 photons, and fitted with emission model \texttt{zphabs*(powerlaw+zgauss)+apec} with the line width set free and the thermal gas abundance frozen at the solar value (see Model (iii) in Table\,\ref{tab:spec} for the resulting best-fit model parameters).}
 \label{fig:spec}
\end{figure*}

Likewise, no prominent extended structures can be seen on the hardness ratio HR=(H-M)/(H+M) map presented in Figure\,\ref{fig:HR}, where we utilized solely the medium (M) and hard (H) bands. This reinforces our conclusion that the 6.0--7.0\,keV signal comes predominantly from the unresolved core of the system, and therefore that the iron fluorescence we see (see the following section) is most likely indicative of the X-ray reflection from a cold neutral absorber in the active nucleus of NGC\,3894. At the same time we emphasize that, in the analzyed {\it Chandra} dataset, the target is unresolved down to a few tens of parsecs, which is the scale of the extension in the Fe K$\alpha$ emission observed in several nearby Compton-thick sources \citep{Marinucci13,Marinucci17,Fabbiano18a,Fabbiano18b}.

\subsection{Spectral Modelling}
\label{sec:spectrum}

Based on the analysis of the X-ray surface brightness profiles in NGC\,3894 presented in the previous Section\,\ref{sec:profile} (see Figure\,\ref{fig:SB}), for the spectral modelling of the active nucleus in the system, we have selected a circular region with the radius of 10\,px centred on the core. For this source extraction region, the background was chosen as an annulus with the inner radius of 20\,px and the outer radius of 40\,px (with the prominent point sources removed -- see Figure\,\ref{fig:chandra}).

\subsubsection{Fitting the Binned Data}

The $0.5-7.0$\,keV spectra for the selected source region and the background, with the corresponding calibration files (arf and rmf), were extracted with the CIAO script \texttt{spec-extract}. The fitting was performed in \texttt{XSPEC} on the background-subtracted binned source spectrum (group minimum 5), including the Galactic absorption with the hydrogen column density $N_{\rm H,\,Gal} = 1.83 \times 10^{20}$\,cm$^{-2}$ \citep{GalNH}. 

\begin{figure*}[!th]
\centering
\includegraphics[width=0.41\textwidth]{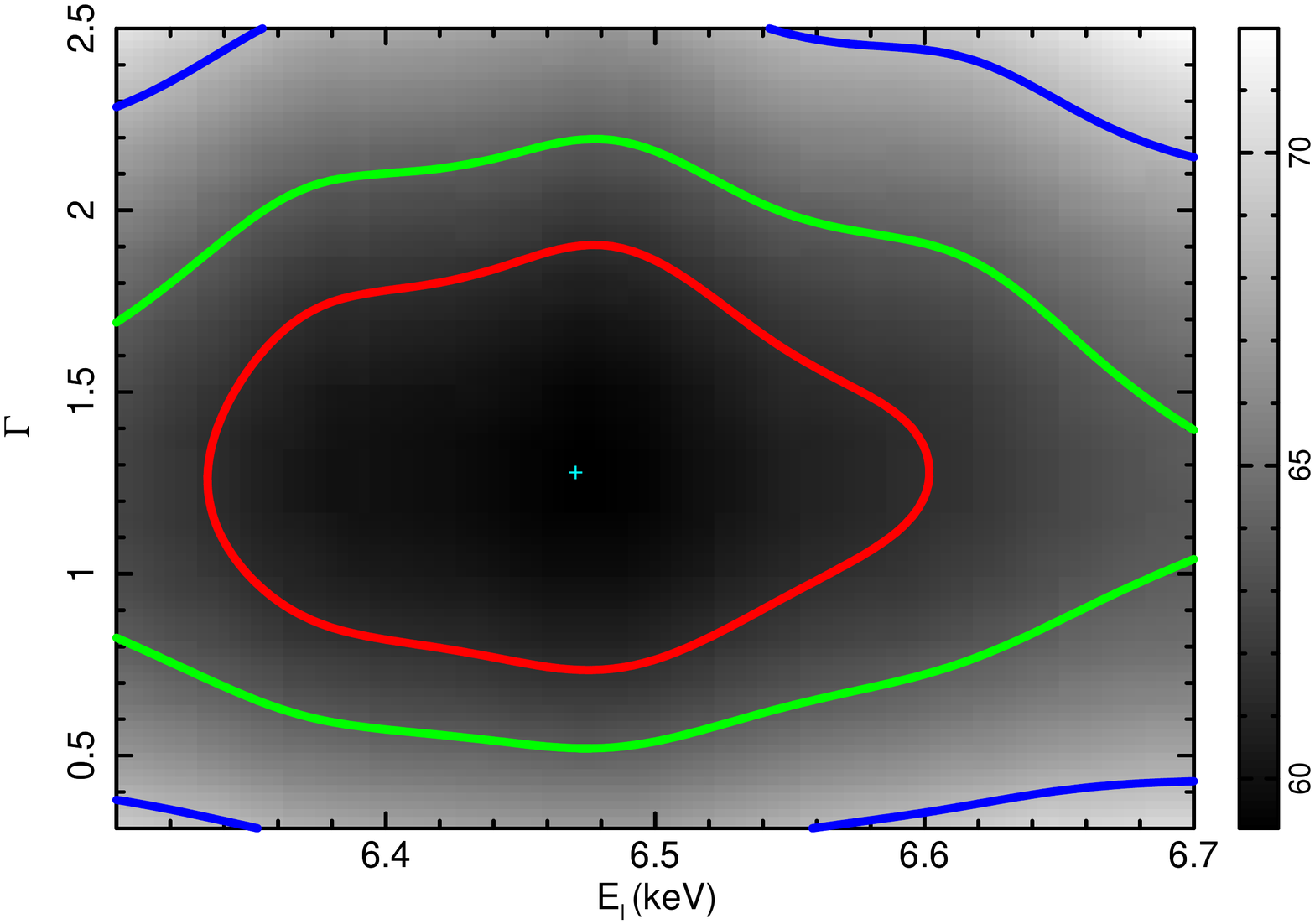}
\includegraphics[width=0.43\textwidth]{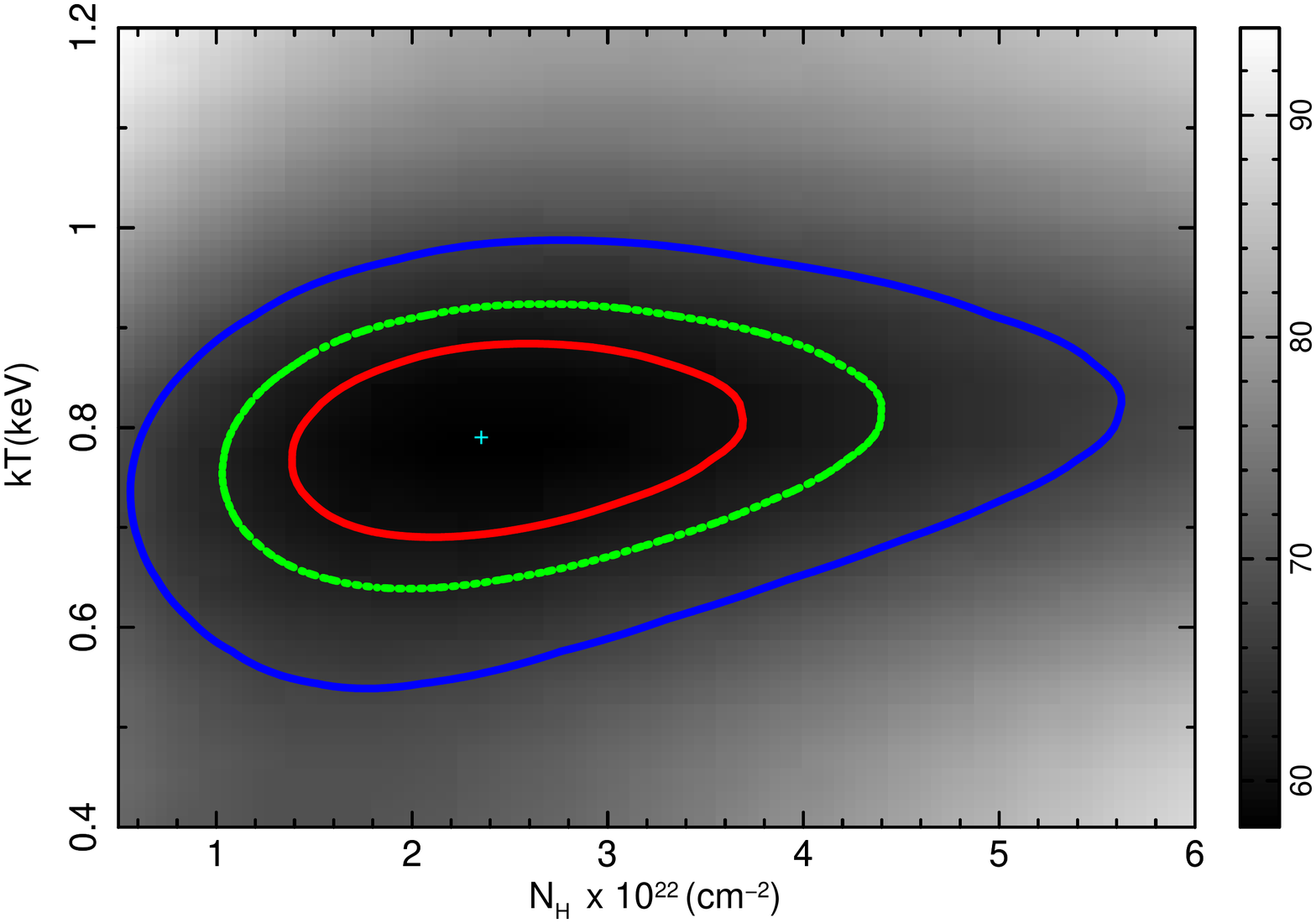}
\includegraphics[width=0.41\textwidth]{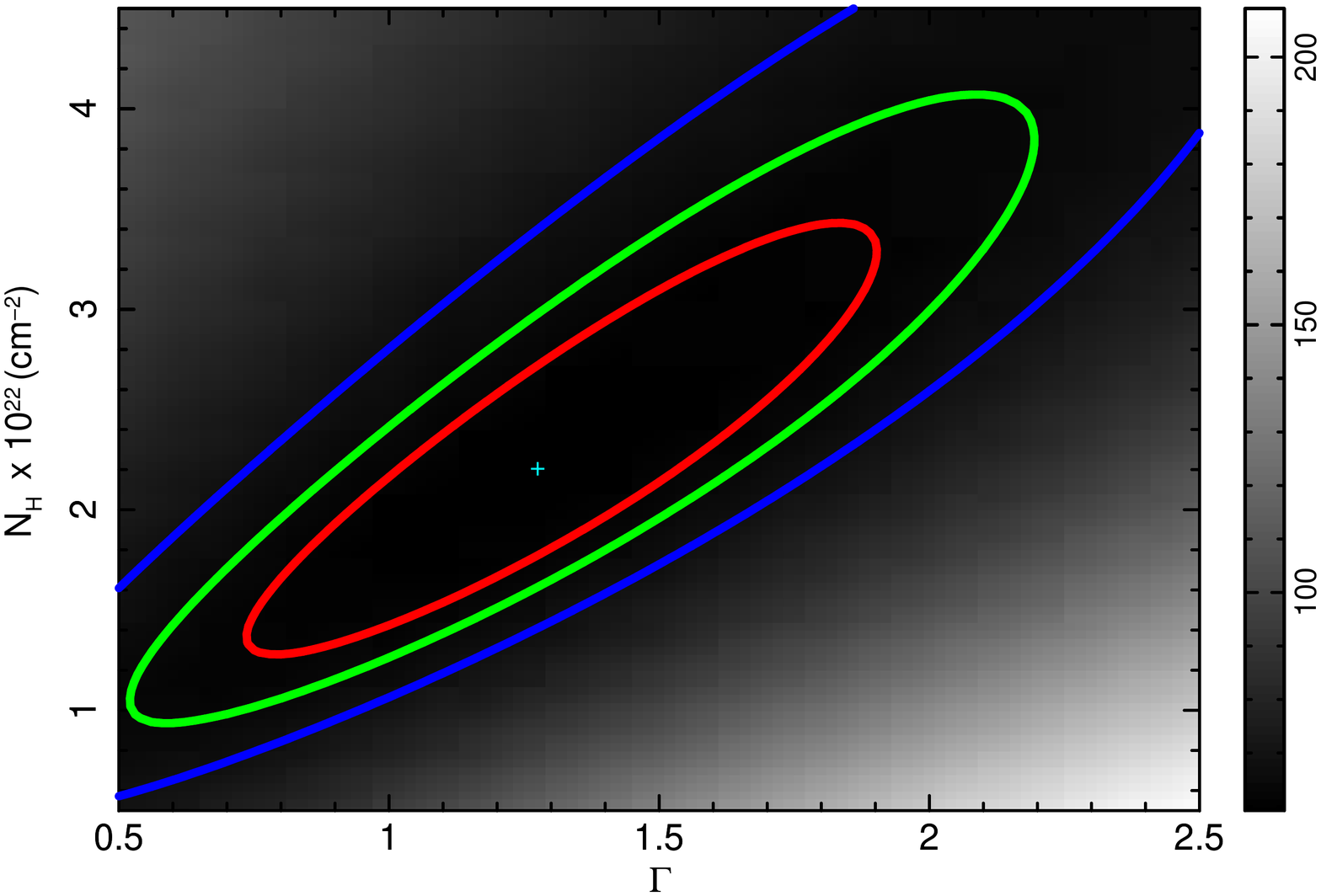}
\includegraphics[width=0.41\textwidth]{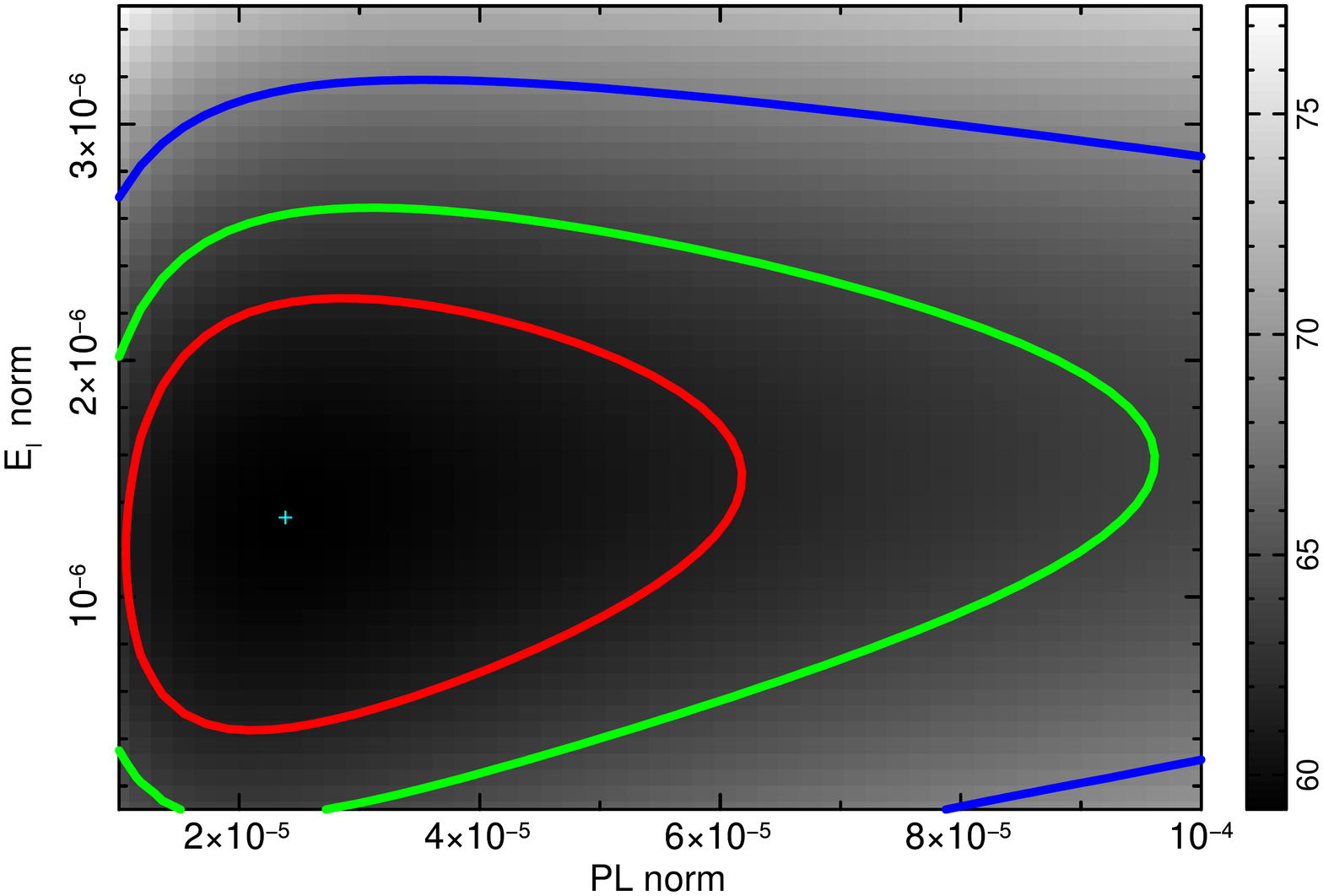}
\caption{Confidence contours of $1\sigma$, $2\sigma$, and $3\sigma$ on the two-thaw-parameters plane for the model (iii) applied to the 0.5--7.0\,keV Chandra spectrum of NGC\,3894, calculated using the Levenberg--Marquardt optimization method: line energy $E_{\ell}$ vs. power-law photon index $\Gamma$ (upper left panel), hydrogen column density $N_{\rm H}$ vs. temperature of the thermal component $kT$ (upper right), power-law photon index $\Gamma$ vs. hydrogen column density $N_{\rm H}$ (lower left), and normalization of the power-law component vs. line normalization (lower right). The grayscale in all the panels corresponds to the final fit statistic values.}
 \label{fig:confcont}
\end{figure*}

Even though the source spectrum appears relatively hard, the simplest model consisting of an absorbed power-law emission component, \texttt{zphabs*powerlaw}, provided a very poor fit to the data. The fit could, on the other hand, be improved significantly by including an additional component representing a collisionally ionized thermal plasma (with the metalicity frozen at the solar value), \texttt{zphabs*powerlaw+apec}, hereafter referred to as model (i). Still, the best-fit slope of the power-law continuum turned out unusually hard. Moreover, positive residuals indicated the presence of emission lines, including a line around 6.4\,keV, consistent with the neutral iron K$\alpha$ emission, and a feature around 4.6\,keV, most likely representing instrumental artefact \citep[in particular, the Si escape peak following the Fe 6.4\,keV emission; see][]{Grimm2009}. 

As the next step, we have therefore added a simple redshifted Gaussian line component to the model, \texttt{zphabs*(powerlaw+zgauss)+apec}, obtaining a satisfactory fit to the data with the line width $\sigma_{\ell}$ either (ii) frozen at 10\,eV, or (iii) kept as a free model parameter. The best-fit values of the (i)-(iii) model parameters are summarized in Table\,\ref{tab:spec}; the background-subtracted source spectrum fitted with latter model (iii), is shown in Figure\,\ref{fig:spec}.

As follows, the thermal emission dominates the radiative output of the source in the soft {\it Chandra} band, i.e. below 1.5\,keV. This component is well modelled as a collisionally ionized plasma (\texttt{apec}) with solar abundance and 0.8\,keV temperature, although we note at the same time some lower-significance positive residuals, which could signal the presence of a photoionized plasma component in addition to the collisionally ionized one, similarly as seen in the another extremely compact CSO with the LINER-type nucleus detected in HE $\gamma$-rays, PKS\,1718-649 \citep[see the discussion in][]{Beuchert2018}. A limited photon statistics combined with a limited {\it Chandra} energy range extending down to only 0.5\,keV, precludes us from a more detailed investigation of this possibility.

Meanwhile, the medium and hard {\it Chandra} bands (1.5--7.0\,keV) are dominated by a non-thermal emission, well approximated by a power-law model with a prominent iron line at 6.4--6.5\,keV, seen through the cold gas with the equivalent hydrogen column density $N_{\rm H} = (2.4 \pm 0.7) \times 10^{22}$\,cm$^{-2}$. The 0.5--7.0\,keV unabsorbed flux of the power-law component reads as $F_{\rm 0.5-7.0\,keV} \simeq 2\times 10^{-13}$\,erg\,cm$^{-2}$\,s$^{-1}$, corresponding to the isotropic luminosity of $L_{\rm X} \simeq 6\times 10^{40}$\,erg\,s$^{-1}$. 

In order to compare the models (i)-(iii), we apply the F-test, obtaining the $p$-value 0.05669 for the models (i) and (ii), 0.21572 for the models (ii) and (iii), and finally 0.06425 for the models (i) and (iii). Hence, we conclude that the \texttt{zphabs*(powerlaw+zgauss)+apec} models (ii) and (iii), which include in addition the redshifted iron line with the width $\sigma$ either frozen at 0.01\,keV or left free, respectively, fit the data better than the \texttt{zphabs*powerlaw+apec} model (i), and as such are preferred at the significance level of $\simeq 94\%$.

In Figure\,\ref{fig:confcont} we present the confidence contours corresponding to the model (iii) parameters. As shown, the probability distribution for the photon index of the power-law component, $\Gamma$, is correlated with the possible values of the intrinsic column density $N_{\rm H}$, and so with the normalization of the PL component. Keeping this in mind, the $1\sigma$ uncertainty range for the EW of the iron line in the model, 
\begin{equation}
    {\rm EW} = \frac{F_{\ell}}{F_{\rm PL}\!(E_{\ell})} \, = \frac{\rm Line \,norm.}{{\rm PL\,norm.} \times (E_{\ell}/{\rm keV})^{-\Gamma}} \, ,
\end{equation}
where $F_{\ell}$ is the total photon flux in the line, and $F_{\rm PL}\!(E_{\ell})$ is the photon flux density of the power-law component at the position of the line, could be estimated as $\sim (0.5-1.9)$\,keV using 200 trials, while for the model (ii) it would read as 0.4--1.1\,keV.

We have also investigate the scenario with a warm ionized absorber (\texttt{warmabs}) replacing the cold neutral absorber (\texttt{zphabs}), but this did not yield any acceptable fits to the data. The fit could not be improved either by replacing the phenomenological model \texttt{powerlaw+zgauss} with a more self-consistent description of a reflection component (e.g., \texttt{pexmon}), but that could be only due to the limited {\it Chandra} photon energy range extending up to only 7\,keV, and as such precluding from any precise characterization of a Compton hump component.

\begin{deluxetable*}{cccccccccc}[!th] 
\tabletypesize{\scriptsize}
\tablecaption{Spectral modelling of the un-binned nuclear spectrum of NGC\,3894 together with the background}
\label{tab:spec-unbinned}
\tablewidth{0pt}
\tablehead{
\colhead{Model} & \colhead{$N_{\rm H}$}  & \colhead{$\Gamma$} & \colhead{PL norm.} & \colhead{$E_{\ell}$} & \colhead{Line norm.} & \colhead{4.6\,keV Line norm.} & \colhead{$\Gamma_{\rm bck}$} & \colhead{Bck PL norm.}  & \colhead{C-stat/DOF} \\
\colhead{~~} & \colhead{$\times 10^{22}$\,cm$^{-2}$} & \colhead{~~} & \colhead{$\times 10^{-6}$} & \colhead{keV} &  \colhead{$\times 10^{-6}$} & \colhead{$\times 10^{-6}$} & \colhead{~~} & \colhead{$\times 10^{-6}$} & \colhead{~~} \\
\colhead{(1)} & \colhead{(2)} & \colhead{(3)} & \colhead{(4)} & \colhead{(5)} & \colhead{(6)} & \colhead{(7)} & \colhead{(8)} & \colhead{(9)} & \colhead{(10)} 
}
\startdata
(a) & $1.2\pm 0.6$ & $0.7\pm 0.3$ & $13_{-5}^{+9}$ & --- & --- & --- & $0.7\pm 0.2$ & $3.1_{-0.7}^{+0.8}$ & 689.21/751\\
(b) & $1.9 \pm 0.7$ & $1.2\pm 0.4$ & $25_{-11}^{+20}$ & $6.5\pm 0.1$ & $2.0_{-0.8}^{+0.9}$ & --- & $0.7\pm 0.2$ & $3.1_{-0.7}^{+0.8}$ & 681.665/749\\
(c) & $2.1\pm 0.7$ & $1.5\pm 0.4$ & $33_{-15}^{+29}$ & $6.5\pm 0.1$ & $2.3_{-0.8}^{+0.9}$ & $1.1\pm0.5$ & $0.7\pm 0.2$ & $3.1_{-0.7}^{+0.8}$ & 672.87/748\\
\enddata
\tablecomments{Col(1) --- model fit; Col(2) --- intrinsic hydrogen column density; Col(3--4) --- photon index and normalization of the power-law emission component; Col(5--6) --- Gaussian line energy and normalization; Col(7) --- 4.6\,keV Gaussian line normalization; Col(8--9) --- photon index and normalization of the power-law emission representing the background; Col(10) --- Cash fitting statistics/DOF. The iron line equivalent width ($1\sigma$ range) reads as EW\,$\simeq 0.8\,(0.5-1.6)$\,keV for the fit (b) with no 4.6\,keV line, and EW\,$\simeq 1.1\,(0.6-1.7)$\,keV for the fit (c) including 4.6\,keV line.}
\end{deluxetable*}

Finally, we note that the same {\it Chandra} dataset for NGC\,3894, has been analyzed before by \citet{She17}, in their investigation of hundreds of nearby galaxies (distances up to 50\,Mpc) with available ACIS pointings. The analysis differ from ours with respect to the source and background extraction regions; the analysis results were however comparable, except of no iron line detected in the analysis by \citeauthor{She17}. In the Appendix\,\ref{A:alternative} below, we discuss this issue in more detail, arguing that with a small source extraction region and large binning of the source spectrum, the iron line could be overlooked indeed.

\subsubsection{Fitting the Un-binned Data}

In the case of low-count spectra, as the one we are dealing with here, spectral fittings performed on the binned data with the $\chi^2$ statistics, could lead to biased results regarding the best-fit model parameters and a significance of the line detection. We have therefore repeated the spectral analysis with \texttt{Sherpa}, this time on un-binned data using the Cash statistics (and the Nelder-Mead optimalization method as before), fitting simultaneously the source and the background regions. The problem we have encountered, however, was that the background spectrum --- in particular its lower-energy segment below 1\,keV --- could not be fitted well in this approach, with any of several models selected. For this reason, the un-binned data fitting was performed exclusively within the 1.5--7.0\,keV photon energy range, where all the ISM-related absorption effects, as well as contributions from the extended thermal components, are expected to be negligible.

\begin{figure}[!th]
\centering
\includegraphics[width=\columnwidth]{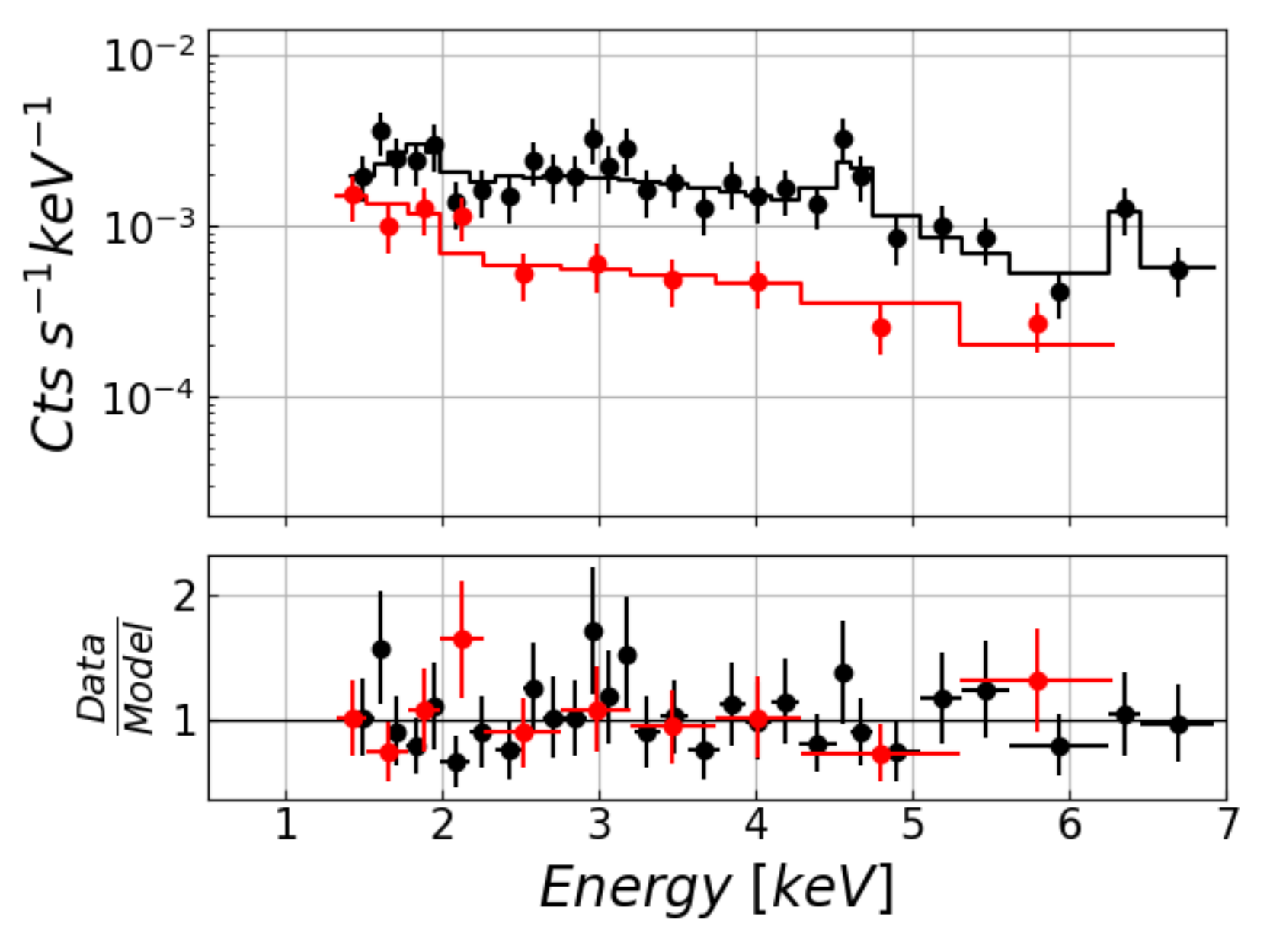}
\caption{$1.5-7$\,keV {\it Chandra} spectrum of the unresolved X-ray core in NGC\,3894 (black symbols), and the background (red symbols), along with the model curves following from simultaneous fitting \texttt{zphabs*(powerlaw+zgauss)+gaussian} performed on the un-binned data using Cash statistics (see Model (c) in Table\,\ref{tab:spec-unbinned} for the resulting best-fit model parameters).}
 \label{fig:spec-unbinned}
\end{figure}

For the background, we have assumed a single unabsorbed power-law model. For the source, we have assume either (a) an absorbed single power-law model \texttt{zphabs*powerlaw}, (b) an absorbed single power-law model with a redshifted Gaussian line, \texttt{zphabs*(powerlaw+zgauss)}, and (c) an absorbed single power-law with a redshifted Gaussian line plus an another Gaussian with the line energy frozen at 4.6\,keV (in order to account for the instrumental feature due to the Si escape),  \texttt{zphabs*(powerlaw+zgauss)+gaussian}. The widths of the lines were frozen at $\sigma =0.1$\,keV. The results of the fitting are summarized in Table\,\ref{tab:spec-unbinned}, and for the model (c) also presented in Figure\,\ref{fig:spec-unbinned}.

As follows, the results of he modelling are in a very good agreement with the ones presented in the previous section ($\chi^2$ fitting of the binned spectrum), returning in particular the equivalent width of the iron line, with the best-fit position $6.5$\,keV, as large as EW\,$\simeq 0.8\,(0.5-1.6)$\,keV for the model (b), and even $1.1\,(0.6-1.7)$\,keV for the model (c), where the ranges enclosed by brackets correspond to $1\sigma$ errors calculated using 200 trials. Moreover, since the change in Cash statistics is distributed approximately as $\Delta \chi^2$, for the model comparison test here we chose the Maximum Likelihood Ratio test, which returns the $p$-values 0.02352 for models (a) and (b), 0.00301 for the models (b) and (c), and finally 0.00098 for the models (a) and (c). This implies that the model (c), despite being the most complex one, best describes the data, and should be selected over the models (a) and (b) with high significance.

\subsection{Broad-band SED}

\begin{figure*}[!th]
\centering
\includegraphics[width=0.9\textwidth]{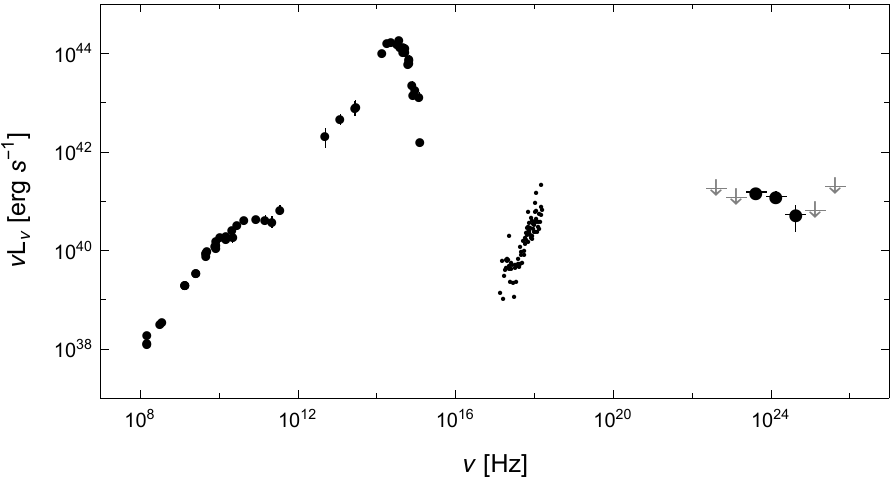}
\caption{The background-subtracted 0.5-7.0\,keV {\it Chandra} spectrum of the central parts of NGC\,3894, along with the {\it Fermi}-LAT spectral datapoints following from the analysis by \citet{Principe2020}, as well as historical radio--to--UV flux measurements for the target collected from the literature (see Appendix\,\ref{A:SED} below). No extinction correction has been applied to the optical/UV datapoints; likewise, the X-ray datapoints represent the observed fluxes, not corrected for the absorption.}
\textit{}
 \label{fig:SED}
\end{figure*}

The background-subtracted 0.5-7.0\,keV {\it Chandra} spectrum of the central parts of NGC\,3894 ($5^{\prime\prime} \simeq 1.2$\,kpc source extraction region), along with the {\it Fermi}-LAT spectral datapoints following from the analysis by \citet{Principe2020}, as well as historical radio--to--UV flux measurements collected from the literature (see Appendix\,\ref{A:SED} below), are presented in the updated broad-band SED Figure\,\ref{fig:SED}. No extinction correction has been applied to the optical datapoints. As shown, the radiative output of the system is dominated by the starlight of the host at near-infrared to optical frequencies \citep[$H$-band absolute magnitude of $-21.80$; see][]{perlman01}, with a likely contribution from the circumnuclear hot dust in the mid-IR range. We note in this context that, in the WISE color-color diagram, our target lies outside of the \citet{Stern12} or \citet{Mateos12} ``Active Galactic Nucleus (AGN) cuts'' \citep[see][]{Kosmaczewski2020}, but at the same time that the efficiency
of the mid-IR colour selections strongly decreases for low-luminosity AGN \citep[see the related discussion in][]{Stern12,Boorman18}.

The non-thermal radiative output of the active nucleus in NGC\,3894, on the other hand, reaches isotropic luminosities $\lesssim 10^{41}$\,erg\,s$^{-1}$ in the radio--to--mm/sub-mm range, and $\lesssim 10^{42}$\,erg\,s$^{-1}$ at HE $\gamma$-ray frequencies. 

\section{Discussion}

\subsection{The Iron Line Fluorescence}

The best-fit line energy for the iron fluorescence detected in the {\it Chandra} data for NGC\,3894, $E_{\ell} = 6.5\pm 0.1$\,keV (see \S\,\ref{sec:spectrum}), is consistent with the position of a neutral K$\alpha$ line, 6.4\,keV, within the errors, although a slight blue-shift could suggest a blend with ionized iron features with much lower intensity. Iron fluorescence seems restricted to the unresolved core, which however corresponds in our case to the inner kpc radius around the galactic nucleus (see \S\,\ref{sec:profile}).

Till now, neutral iron K$\alpha$ line has been found in the two other young radio galaxies of the CSO type, namely 1404+286 \citep[aka OQ+208;][]{Guainazzi04,Sobolewska19b}, and 2021+614 \citep{Sobolewska19a}, none of which is however detected at HE $\gamma$-rays with {\it Fermi}-LAT.

We also note that iron fluorescence features in the X-ray spectra of the three other CSOs 1934--638, 1946+708, and 1607+268, claimed based on the Beppo-SAX or first XMM-{\it Newton} observations \citep{Risaliti03,Tengstrand09}, have not been confirmed with the {\it Chandra} or the following XMM-{\it Newton} data \citep{Siemiginowska16,Sobolewska19a}. On the other hand, yet another CSO 0710+439 has been found to display a very strong ionized iron line at $E_{\ell} \simeq 6.6$\,keV, with large EW\,$\sim 1$\,keV \citep{Siemiginowska16}. Ionized iron line at $\simeq 6.7$\,keV with the EW\,$\sim 0.3$\,keV has been also found in the two other radio galaxies with young inner structures (although not strictly of the CSO morphological type), namely PMN\,J1603-4904 \citep{Muller15,Goldoni16}, and CGCG\,292--057 \citep{Balasubramaniam20}.

When comparing NGC\,3894 with 1404+286 and 2021+614 \citep[see Table\,1 in][and references therein]{Kosmaczewski2020}, we note that all the three sources are similarly compact (linear sizes of $\sim 4$\,pc, $\sim 10$\,pc and $\sim 16$\,pc, respectively), but the latter two are (a) much more X-ray luminous (by approximately two and three orders of magnitude, respectively), and moreover (b) Compton-thick, i.e. characterized by the equivalent hydrogen column densities $N_{\rm H} \sim 10^{24}$\,cm$^{-2}$, which is about two order of magnitude larger than $N_{\rm H}$ inferred in our analysis for NGC\,3894. At the same time, the equivalenth width of iron K$\alpha$ line in 1404+286 and 2021+614 is lower, EW\,$\sim 0.6$\,keV and $\sim 0.2$\,keV, respectively. 

It may therefore be that, in our relatively simple modelling of the {\it Chandra} spectrum with a limited energy range and limited photon statistics, we seriously under-estimate the true absorbing column density (and hence also the intrinsic X-ray luminosity), for example due to the complex structure of the circumnuclear cold neutral gas and dust in the source. And indeed, the ``narrow'' neutral iron lines with equivalent widths of the order of 1\,keV, are expected rather for Compton-thick AGN, while for the objects with $N_{\rm H} < 10^{23}$\,cm$^{-2}$ the iron K$\alpha$ EWs are typically of the order of 100\,eV only \citep[e.g.,][]{Fukazawa11}. Investigating this possibility in more detail, awaits however deeper X-ray exposures of our target with the instruments such as XMM-{\it Newton} and Nu-STAR (but see also the discussion in the following sub-section).

In this context, we note that the result of our analysis regarding the EW value are in a accord with the ``X-ray Baldwin effect'', or the ``Iwasawa--Taniguchi effect'' established for Seyfert galaxies of both type 1 and 2 \citep{Iwasawa93}, which states an anti-correlation between the EW of the Fe K$\alpha$ emission line at 6.4\,keV, and the X-ray continuum luminosity \citep[see, e.g.,][and references therein]{Page04,Fukazawa11,Ricci14,Boorman18}.

In particular, we refer to the work by \citet{Boorman18}, who analyzed exclusively the heavily obscured (Compton-thick) AGN, i.e. the sources for which the intrinsic (i.e., un-absorbed) X-ray fluxes at $<10$\,keV photon energies are difficult to assess. For those, instead of the X-ray luminosities, \citeauthor{Boorman18} used the mid-IR 12\,$\mu$m luminosities based on the WISE and {\it Spitzer} Multiband Imaging Photometer
(MIPS) data, obtaining a statistically significant correlation
\begin{eqnarray}
\log\left(\frac{\rm EW_{\rm Fe\,K\alpha}}{\rm eV}\right) & = & (2.87 \pm 0.05)   - \\
&& (0.08 \pm 0.04) \times \log\left(\frac{L_{\rm 12\,\mu m}}{10^{44}\,{\rm erg\,s^{-1}}}\right)\, . \nonumber 
\end{eqnarray}
For the WISE 12\,$\mu$m luminosity of NGC\,3894, $L_{\rm 12\,\mu m} \simeq 1.0 \times 10^{42}$\,erg\,s$^{-1}$ \citep{Kosmaczewski2020}, this correlation would therefore imply EW\,$= 1.07^{+0.37}_{-0.28}$\,keV, which is in a (surprisingly) very good agreement with the best-fit value following from our spectral analysis.

\subsection{X-ray and Radio Absorbers}

Our spectral analysis reveal that the nuclear power-law X-ray emission component is seen through the cold absorbing material with the equivalent hydrogen column density $N_{\rm H} \simeq 2.5 \times 10^{22}$\,cm$^{-2}$. This best-fit value turns out about $\sim 2$\,dex larger than the \emph{average} value for the neutral hydrogen column density $N_{\rm HI}$ inferred from the HI absorption features, assuming the spin temperature of the HI gas of 100\,K, and a covering factor of the order of unity \citep{peck98,perlman01,gupta06,emonts10}. Interestingly, this would be then in agreement with the linear relation claimed by \citet[see section 6.1, sample $D^{\prime}$ therein]{Ostorero17}, namely
\begin{equation}
    \log N_{\rm HI} = 6.7^{+4.8}_{-4.2} + 0.63^{+0.19}_{-0.22} \times \log N_{\rm H} \, ,
    \label{eq:luisa}
\end{equation}
with the intrinsic spread $\epsilon_{\rm H} = 0.92^{+0.30}_{-0.21}$, which --- for our best-fit value $N_{\rm H}$ --- implies that $N_{\rm HI}$ should fall (with $68\%$ probability) within a factor of about eight from $\sim 6.5 \times 10^{20}$\,cm$^{-2}$. This agreement reinforces the main conclusions by \citeauthor{Ostorero17} that, in CSOs, ``the X-ray and radio absorbers are either co-spatial or different components of a continuous structure''.

And conversely, if the absorbing gas column density is significantly under-estimated in our {\it Chandra} spectral modelling, as speculated in the previous sub-section, the average $N_{\rm HI}$ value according to the \citeauthor{Ostorero17} correlation, should be larger than observed. For example, assuming a Compton-thick limit of $N_{\rm H} = 1.5 \times 10^{24}$\,cm$^{-2}$, from equation\,\ref{eq:luisa} we obtain the expected $N_{\rm HI} \simeq 8.5 \times 10^{21}$\,cm$^{-2}$. Still, we emphasize a rather large intrinsic spread in the correlation \ref{eq:luisa} (factor of about eight), interpreted as due to a spread in the HI gas spin temperature \citep[which may vary substantially from the typically assumed ``universal'' value of 100\,K; see the discussion in][]{Ostorero17}.

\subsection{The Overall Energetic of the Active Nucleus}

The optical velocity dispersion of the central $3.5^{\prime\prime}$ of the galaxy was measured by \citet{van15}, with the 10\,m Hobby--Eberly Telescope (HET) at McDonald Observatory, as $\sigma = 326.0  \pm 10.1$\,km\,s$^{-1}$. \citeauthor{van15} provided also the best-fit relation between the velocity dispersion and the black hole mass
\begin{equation}
\log\left(\frac{\mathcal{M}_{\rm BH}}{M_{\odot}}\right) = \alpha + \beta \times \log\left(\frac{\sigma}{200\,{\rm  km\,s^{-1}}}\right) \, ,
\end{equation}
with $\alpha = 8.24\pm 0.06$, $\beta = 5.28\pm 0.37$, and the intrinsic scatter $\epsilon_{\mathcal{M}} = 0.40\pm 0.21$, consistent with \citet{McConnell13}. With such, for NGC\,3894 we obtain $\mathcal{M}_{\rm BH} \sim 2 \times 10^9 M_{\odot}$ withn a factor of 2.5, which is in a very good agreement with the value obtained by \citet{wil10}, based on the [Oiv] 25.8\,$\mu$m transition line. The estimate does not change if the $\mathcal{M}_{\rm BH} - \sigma$ scaling relation for ellipticals and bulge-like systems from \citet{Graham11} is used instead ($\alpha = 8.13\pm 0.05$, $\beta = 5.13\pm 0.34$). We note that, on the other hand, the estimate based on the bulge--luminosity relation using the $g$ and $r$ photometry from the Sloan Digital Sky Survey (SDSS), returns the black hole mass lower by one order of magnitude \citep[see][]{wil10}.

With $\mathcal{M}_{\rm BH} \sim 2 \times 10^9 M_{\odot}$, the corresponding Eddington luminosity reads as $L_{\rm Edd} \sim 3 \times 10^{47}$\,erg\,s$^{-1}$, and the X-ray radiative output of the NGC\,3894 unresolved core, in the Eddington units, as $L_{\rm X}/L_{\rm Edd} \sim 2 \times 10^{-7}$. Likewise, the jet/lobe-related HE $\gamma$-ray emission of the source, as estimated by \citet{Principe2020}, turns out to be one order of magnitude larger than that, namely $L_{\gamma}/L_{\rm Edd} \sim 2 \times 10^{-6}$. 

Unfortunately, the bolometric accretion-related luminosity in the target, $L_{\rm bol}$, is unknown and hard to assess with the available optical spectroscopy of the nuclear region \citep[see][]{kim89,Goncalves04}. For a rough estimate, one could try to use the observed X-ray core emission, assuming that it provides a limit on the disk coronal radiative output, and hence on the direct disk continuum emission. However, the ratio $L_{\rm X}/L_{\rm bol}$ is known to vary widely among the youngest radio galaxies of the CSO type, from $\gtrsim 0.1$ down to $<10^{-3}$ \citep[see][]{Wojtowicz20}.

Alternatively, one may use the observed IR flux, assuming that it provides a limit on the emission of the circumnuclear dust in the source. Hot dusty tori in active galaxies are, in general, expected to reprocess about $\eta_{\rm DT} \sim 10\%$ of the accretion disk bolometric luminosity. With such, the integrated $8-1000$\,$\mu$m luminosity of NGC\,3894 as estimated by \citet{Kosmaczewski2020}, would then imply an upper limit on the accretion disk luminosity $L_{\rm bol} \sim 5 \times 10^{43}$\,erg\,s$^{-1}$, or in terms of the accretion rate $\lambda_{\rm Edd} \equiv L_{\rm bol}/L_{\rm Edd} \sim 10^{-4}$. We stress, however, that the above estimate is highly uncertain, because of the uncertain value of the radiative efficiency factor $\eta_{\rm DT}$, and also due to the fact that the ``callorimetric'' properties of circumnuclear dust can only reflect the nuclear power output averaged over some longer period of time, rather than the accretion disk luminosity at a given (present) moment.

Finally, we estimate the jet kinetic power $P_{\rm j}$ in the radio source 1146+596. In particular, assuming the minimum power condition, linear size and kinematic age of the radio structure LS\,$\simeq 4$\,pc and $\tau_{\rm j} \sim 60$\,yr \citep{Principe2020}, as well as the VLBA 5\,GHz flux of 387.6\,mJy \citep{Helmboldt07}, corresponding to the monochromatic radio power $L_{\rm 5\,GHz} \sim 0.6 \times 10^{40}$\,erg\,s$^{-1}$, from equation\,4 in \citet{Wojtowicz20} we obtain $P_{\rm j} \sim 2 \times 10^{42}$\,erg\,s$^{-1}$, or $P_{\rm j}/L_{\rm Edd} \sim 0.7 \times 10^{-5}$. This implies that the observed X-ray luminosity (exclusively the power-law emission component), could, in principle, be contributed by the jet emission, as it represents only a small fraction of the jet kinetic power, of the order of a few percent. The HE $\gamma$-ray emission, on the other hand, if due to the jets/compact lobes, would require a much higher radiative efficiency, at the level of $\gtrsim 10\%$. Still one has to keep in mind that the above estimate for $P_{\rm j}$, relies on the assumed minimum power condition, and as such corresponds to rather a lower limit for a true jet kinetic power in the source.

\section{Conclusions}

In this paper we report on our analysis of the archival {\it Chandra} X-ray Observatory data for the central part of the elliptical/S0 galaxy NGC\,3894, hosting a radio source 1146+596, one of the youngest and nearest confirmed CSOs, and one of only several young radio galaxies (of ``non-blazar'' type) detected at HE $\gamma$-rays with the {\it Fermi}-LAT. The analyzed data consist of a single 40\,ksec-long ACIS exposure. We have found that the core spectrum is best fitted by a combination of a collisionally ionized thermal plasma with the temperature of $\simeq 0.8$\,keV, and a moderately absorbed power-law component (photon index $\Gamma = 1.4\pm 0.4$, hydrogen column density $N_{\rm H}/10^{22}$\,cm$^{-2}$\,$= 2.4\pm 0.7$). We have also detected the iron K$\alpha$ line at $6.5\pm 0.1$\,keV, with a relatively large equivalent width of EW\,$= 1.0_{-0.5}^{+0.9}$\,keV. The soft thermal component is undoubtedly extended on the scale of the galaxy host. The iron fluorescence, on the other hand, seems restricted to the unresolved inner kpc radius around the galactic nucleus. The results summarized above, following from fitting the binned data, were supported by the additional analysis performed on the un-binned data, strengthening our claim of the iron line detection.

We conclude that the detected iron line is therefore most likely indicative of the X-ray reflection from a cold neutral absorber in the central regions of NGC\,3894. Yet the emerging relatively large EW of the feature, seems rather confusing, keeping in mind only a moderate obscuration implied by our spectral analysis. For this reason, we speculate that in our relatively simple modelling of the {\it Chandra} data with a limited energy range and limited photon statistics, we could seriously under-estimate the true absorbing column density in the source. This speculation seems supported by the agreement of our best-fit EW Fe\,K$\alpha$ value with the $EW_{\rm Fe\,K\alpha} \propto L_{\rm 12\,\mu m}^{-0.08}$ anticorrelation following from the Iwasawa--Taniguchi (or the ``X-ray Baldwin'') effect, established for heavily obscured AGN by \citet{Boorman18}, for a given 12\,$\mu$m luminosity of the target as measured with WISE. 

A moderate X-ray obscuration in the NGC\,3894 nucleus seems, on the other, supported by the agreement of our best-fit $N_{\rm H}$ value and the average neutral hydrogen column density $N_{\rm HI}$ inferred from the redshifted HI absorption seen against the continuum emission of the central compact radio structure, with the $N_{\rm HI} \propto N_{\rm H}^{+0.63}$ correlation claimed by \citet{Ostorero17}.

We also update on the broad-band SED of NGC\,3894, and discuss the overall energetics of the active nucleus in the system. In particular, based on the optical velocity dispersion of the central $3.5^{\prime\prime}$ of the galaxy measured by \citet{van15}, we estimate the black hole mass in the system as $\mathcal{M}_{\rm BH} \sim 2 \times 10^9 M_{\odot}$ within a factor of 2.5. We also attempt to estimate roughly the accretion rate in the source, based on the observed IR luminosity (considered as a proxy for the radiative output of the circumnuclear hot dusty torus), as $\lambda_{\rm Edd} \sim 10^{-4}$. Such a low accretion rate is in agreement with the LINER classification of the active nucleus in NGC\,3894. Finally, we also estimate the minimum kinetic power of compact jets in 1146+596 as $P_{\rm j} \sim 2 \times 10^{42}$\,erg\,s$^{-1}$.

\begin{acknowledgements}
This work was supported by the Polish NSC grant 2016/22/E/ST9/00061 (K.B., \L .S., V.M., R.T., and D.\L .K), and the Chandra guest investigator program GO5-16109X. This research was supported in part by NASA through contract NAS8-03060 (A.S., M.S.) to the Chandra X-ray Center.
Work by C.C.C. at the Naval Research Laboratory is supported by NASA DPR S-15633-Y. The authors thank G. Principe for providing the {\it Fermi}-LAT fluxes for the source, and the anonymous Referee for very useful comments and suggestions. K.B. acknowledges the help and support from the CXC HelpDesk Team.
\vspace{5mm}
\facilities{Chandra (ACIS)}
\software{CIAO \citep{Fruscione06}, XSPEC \citep{Arnaud1996}, Sherpa \citep{Freeman01}, ChaRT \citep{Carter03}, and MARX \citep{Davis12}}

\end{acknowledgements}

\appendix

\section{Low-Frequency Spectral Measurements} 
\label{A:SED}

The radio source 1146+596 is known to be variable on year timescales. For instance, according to the continuous OVRO monitoring since 2008 \citep{ric11}, the 15\,GHz flux densities vary from $\sim 400$ to 500\,mJy. The few integrated measurements at 8\,GHz indicate similar variations from 436 to 520\,mJy (see Table\,\ref{tab:radio}). From available images in the RFC database\footnote{\url{http://astrogeo.org/rfc/}}, this level of change can be attributed to changes in the VLBI core, with observed 8\,GHz flux densities of 66\,mJy in 1994 Aug, and 155 to 183\,mJy in 2017 Apr and Aug, respectively. The VLBI-scale jet knots appear variable as well, but with typical 5-8\,GHz flux densities observed at levels of only $\sim10$'s mJy \citep{Principe2020}.

Historical IR--optical--UV flux density measurements of NGC\,3894 collected from the literature\footnote{see also \url{http://ned.ipac.caltech.edu}}, with no extinction correction applied, are summarized in Table\,\ref{tab:optical}.

\begin{deluxetable}{ccc}
\tabletypesize{\footnotesize}
\tablecaption{Historical integrated radio flux density measurements of NGC\,3894.}
\label{tab:radio}
\tablewidth{0pt}
\tablehead{
\colhead{$\nu$} & \colhead{$S_{\nu} \pm \Delta S_{\nu}$}  & \colhead{Catalog/Name* \& Reference} \\
\colhead{[GHz]} & \colhead{[mJy]} & \colhead{~~}
}
\startdata
\\
90 & 160 $\pm$ 20 & \citet{pus81} \\
30 & 362 $\pm$ 22 & OCRA-p \citep{low07}\\
22 & 390 $\pm$ 20 & \citet{lee17} \\
15 & 441 $\pm$ 2 & OVRO-40m \citep{ric11}\\
10.7 & 600 $\pm$ 48 & S4 1146+59  \citep{pau78}\\
8.5 & 436 $\pm$ 4 & \citet{fas01} \\
8.4 & 521.7 $\pm$ 52.2 & CRATES J114850.36+592456.4* \citep{hea07} \\ 
8.1 & 520 $\pm$ 52 & NRAO 300-f* \citep{war74}\\
5.0 & 660 $\pm$ 20 & \citet{wro83} \\
4.9 & 606 $\pm$ 9     & S4 1146+59 \citep{pau78}\\
4.85 & 627 $\pm$ 62 & 87GB 114610.9+594135 \citep{gre91} \\ 
2.7 & 440 $\pm$ 11   & S4 1146+59 \citep{pau78}\\
2.7 & 440 $\pm$ 44   & NRAO 300-f* \citep{war74}\\
1.4 & 471.6 $\pm$ 47.2 & FIRST J114850.3+592456* \citep{bec95}\\
1.4 & 481.4 $\pm$ 14.4 & NVSS J114850+592457  \citep{con98}\\ 
0.365 & 327 $\pm$ 19 & TXS 1146+596  \citep{dou96} \\ 
0.325 & 331 $\pm$ 33 & WN B1146.1+5941*  \citep{ren97}\\
0.15 & 282.8 $\pm$ 28.6 & TGSSADR J114850.2+592456  \citep{int17} \\ 
0.151 & 430 $\pm$ 43 & 7C 114610.00+594129.00*  \citep{hal07}\\
0.151 & 300 $\pm$ 30 & 6C 114613+594127*  \citep{hal90}\\
\hline
43.3 & 330 $\pm$ 33 & VLA Calibrator Manual*  \citep{per03} \\ 
22.5 & 280 $\pm$ 60 & VLA Calibrator Manual* \citep{per03}  \\ 
14.9 & 380 $\pm$ 60 & VLA Calibrator Manual* \citep{per03}  \\ 
8.4 & 630 $\pm$ 69 & VLA Calibrator Manual* \citep{per03}   \\ 
4.9 & 520 $\pm$ 78 & VLA Calibrator Manual* \citep{per03}   \\ 
1.4 & 480 $\pm$ 68 & VLA Calibrator Manual* \citep{per03}   \\ 
\\
\enddata
\tablecomments{*Catalogs/Names indicated with asterisks did not quote uncertainties, and $10\%$ was assumed here.}
\end{deluxetable}

\begin{deluxetable}{ccc}
\tablecaption{Historical integrated IR-optical-UV flux density measurements of NGC\,3894.}
\label{tab:optical}
\tablewidth{0pt}
\tablehead{
\colhead{$\nu$} & \colhead{$S_{\nu}$}  & \colhead{$\pm \Delta S_{\nu}$} \\
\colhead{[GHz]} & \colhead{[mJy]} & \colhead{[mJy]}
}
\startdata
\\
1.50E2 & 94.9   & 18.3 \\
2.22E2 & 57.6   & 14.6\\
3.53E2 & 62.9   & 13.3\\
5.00E3 & 140    & 59\\
1.20E4 & 130    & 28\\
2.83E4 & 93     & 29\\
2.97E4 & 93     & 29\\
1.38E5 & 252    & 3.5\\
1.82E5 & 303    & 3.1\\
2.4E5  & 241    & 2.2\\
3.25E5 & 157    & 2.9E--1\\
3.79E5 & 168    & 15.5    \\
3.89E5 & 119    & 1.1E--1\\
4.68E5 & 98.3   & 9.05\\
4.77E5 & 76.3   & 1.41E--1\\
5.42E5 & 81.1   & 11.2    \\
5.42E5 & 66.2   & 4.27\\
6.17E5 & 33.9   & 6.24E--2\\
6.81E5 & 36.8   & 5.45\\
6.81E5 & 31.7   & 3.21\\
6.81E5 & 37.8   & 5.20\\
8.19E5 & 9.33   & 1.30\\
8.36E5 & 5.64   & 2.08E--2\\
9.52E5 & 6.25   & 1.91\\
1.20E6 & 3.66   & 0.78\\
1.29E6 & 0.419  & 5.87E--3\\
\\
\enddata
\end{deluxetable}

\section{Additional Spectral Modelling} 
\label{A:alternative}

\citet{She17} investigated 719 galaxies located within 50\,Mpc, with available {\it Chandra}-ACIS pointings; their sample of objects included NGC\,3894. With a smaller source extraction region and a closer background when compared to our analysis, \citeauthor{She17} binned the background-subtracted source spectrum of the target with minimum 15 photons, and applied the spectral model including a thermal emission component and an absorbed power-law, obtaining comparable results with respect to the temperature of the thermal plasma, $kT \sim 0.85 \,\, (0.20-1.10)$\,keV, the equivalent hydrogen column density, $N_{\rm H}/10^{22}$\,cm$^{-2}$\,$\simeq 2.77 \,\, (1.70-4.14)$, as well as the normalization and slope of the power-law component, $\Gamma=1.23 \,\, (0.65 - 1.90)$. No iron line was detected in the analysis by \citeauthor{She17}.

\begin{figure}[!th]
\centering
\includegraphics[width=\textwidth]{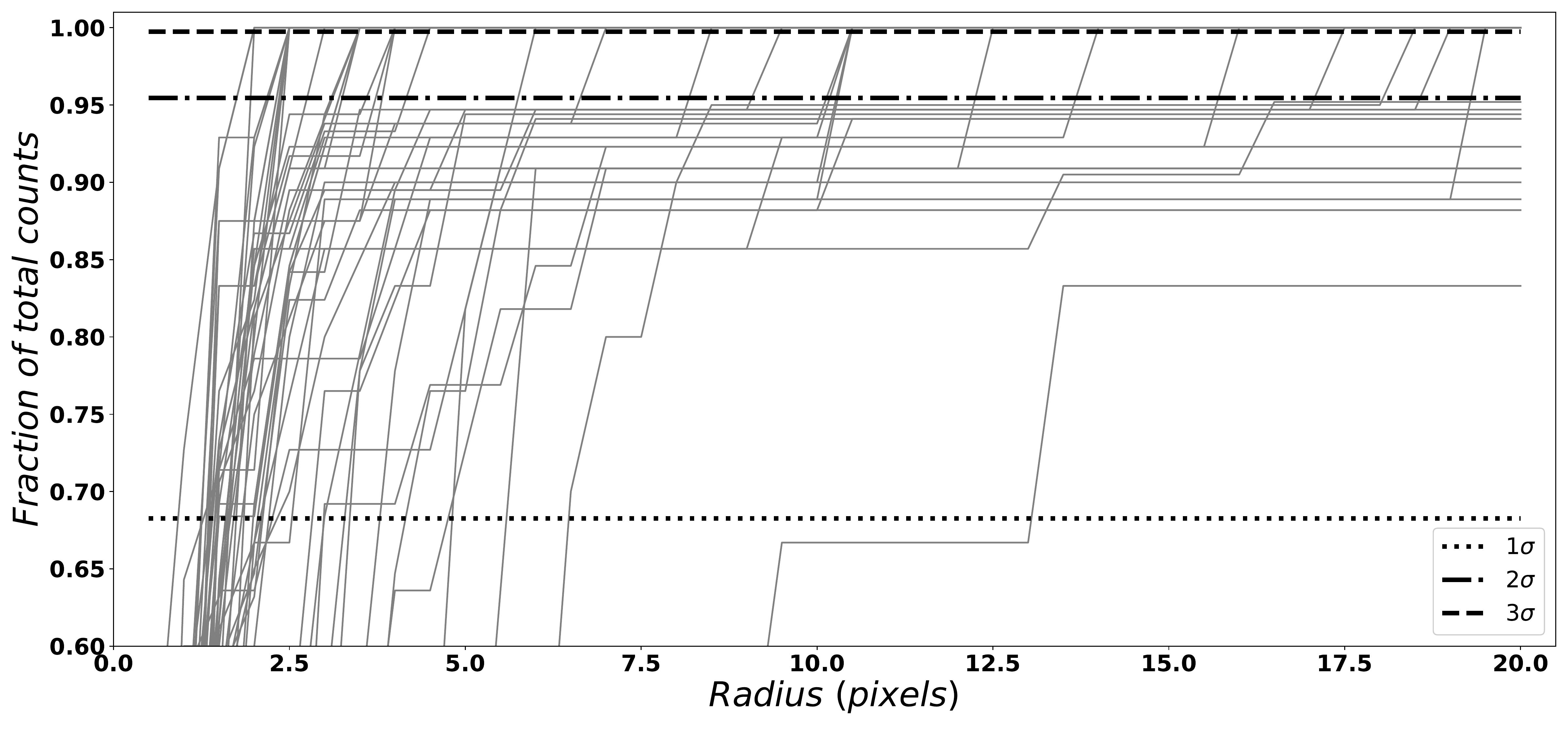}
\caption{Enclosed count fraction for 50 simulated PSF images in the narrow range 6.0-7.0\,keV; horizontal dotted lines from bottom to top correspond to $1\sigma$, $2\sigma$, and $3\sigma$ count fractions, respectively.}
\textit{}
 \label{fig:PSF}
\end{figure}

In order to address the discrepancy regarding a presence of the iron line in the nuclear spectrum of NGC\,3894, we repeat our analysis by choosing a smaller source extraction region of 5\,px radius, and a close background from within $10-20$\,px, consistently with \citet{She17}. First we note, however, that for many realizations of the simulated 6.0--7.0\,keV PSF for the target (see Section\,\ref{sec:profile} above), the 5\,px source extraction region encompasses less than $2\sigma$ of the count fraction, as visualized in Figure\,\ref{fig:PSF}. Indeed, with the 5\,px source extraction region, we measure 22 net counts for the target within the hard band, corresponding to about $85\%$ of the net counts measured with the 10\,px source extraction region.

\begin{figure}[!th]
\centering
\includegraphics[width=0.75\textwidth]{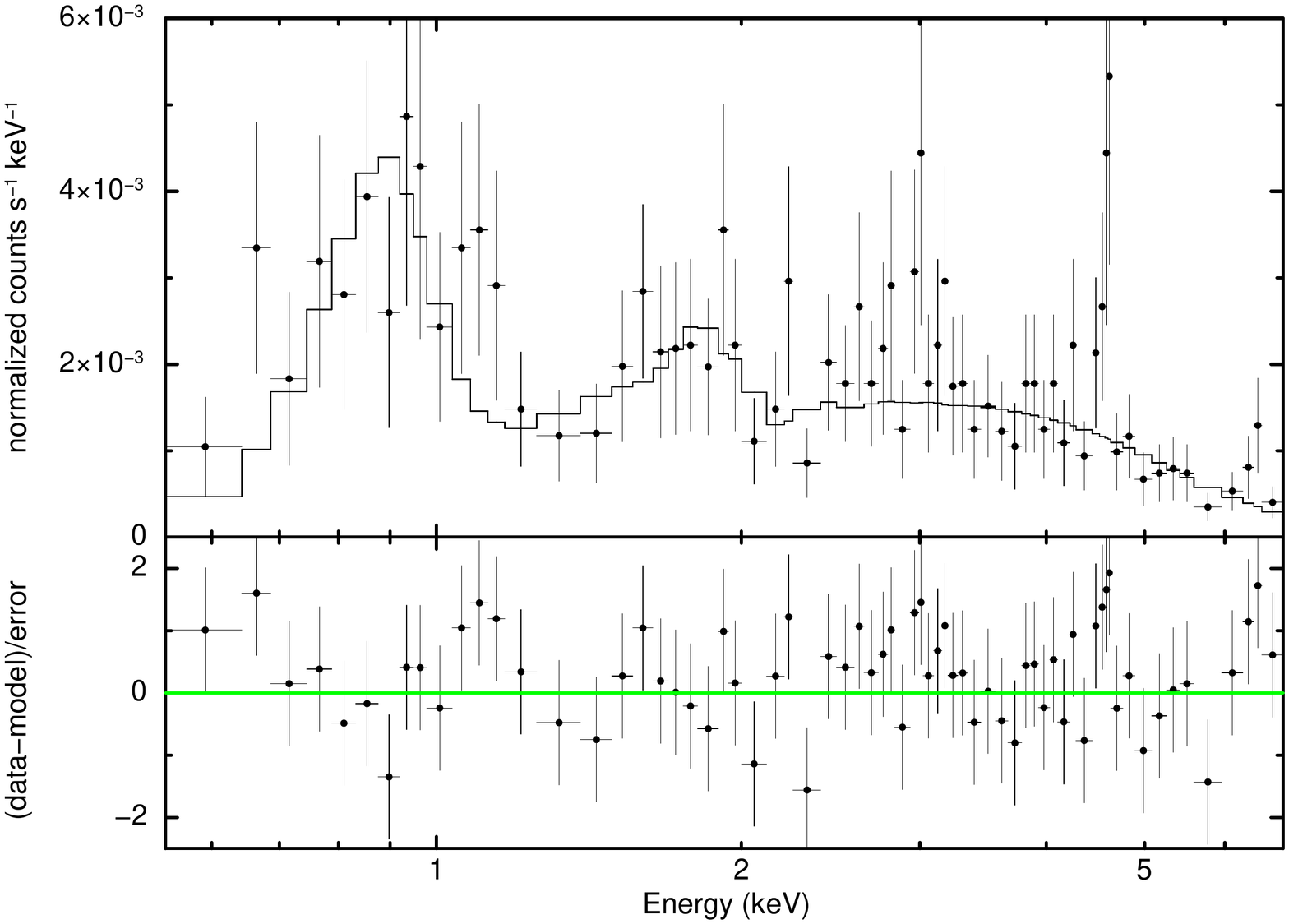}
\includegraphics[width=0.75\textwidth]{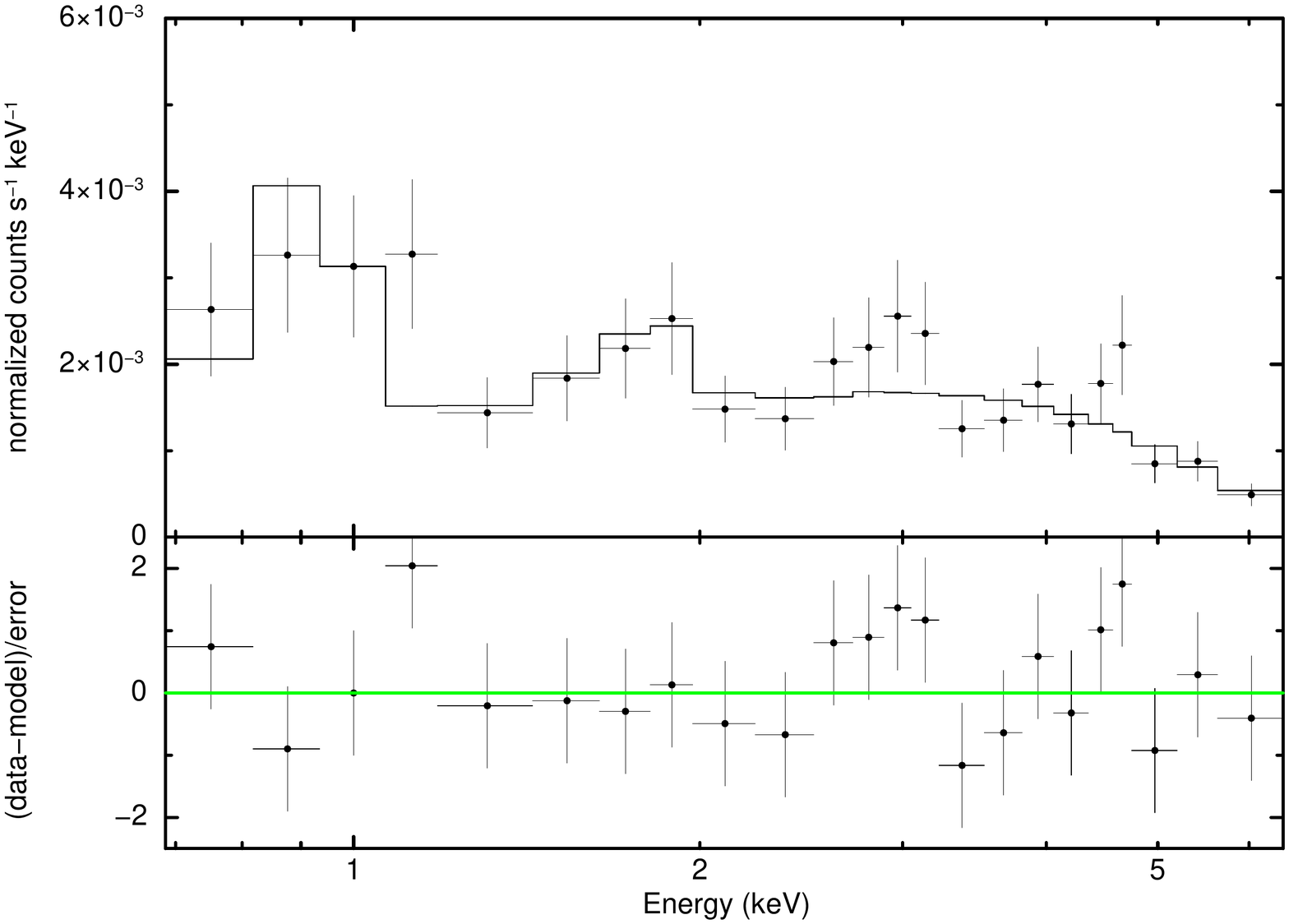}
\caption{Background-subtracted $0.5-7$\,keV {\it Chandra} spectrum of the unresolved X-ray core in NGC\,3894 corresponding to the small source extraction region (radius $=5$\,px) and a close background (annulus with radii $=10-20$\,px), fitted with emission model \texttt{zphabs*powerlaw+apec}, with the applied grouping minimum 5 photons (upper panel) and 15 photons (lower panel). See Table\,\ref{tab:alternative} for the resulting best-fit model parameters.}
\textit{}
 \label{fig:she}
\end{figure}

We next fit the background-subtracted source spectrum, grouped with minimum 5 or 15 photons, assuming a simple model \texttt{zphabs*powerlaw+apec}. The results of the fitting are presented in Figure\,\ref{fig:she}, and summarized in Table\,\ref{tab:alternative}. As follows, for the small source extraction region the iron line can still be noted in the binned spectrum (or in the fitting residuals) but only for a small binning (min 5), while it seemingly disappears for a larger binning (min 15). This, in fact, should be expected keeping in mind the very limited photon statistics in the hard {\it Chandra} band. The source spectrum and the model fit corresponding to grouping with minimum 15 photons, are in agreement with \citet{She17}.

\begin{deluxetable}{ccccccc}
\tabletypesize{\footnotesize}
\tablecaption{Additional spectral modelling of the NGC\,3894 nucleus}
\label{tab:alternative}
\tablewidth{0pt}
\tablehead{
\colhead{Model} & \colhead{$N_{\rm H}$}  & \colhead{$kT$} & \colhead{APEC norm.$^{\dagger}$} & \colhead{$\Gamma$} & \colhead{PL norm.$^{\dagger}$}  & \colhead{$\chi^{2}$/DOF} \\
\colhead{~~} & \colhead{$\times 10^{22}$\,cm$^{-2}$} & \colhead{keV} & \colhead{$\times 10^{-6}$\,\texttt{XSPEC}} & \colhead{~~} & \colhead{$\times 10^{-6}$\,\texttt{XSPEC}} & \colhead{~~} \\
\colhead{(1)} & \colhead{(2)} & \colhead{(3)} & \colhead{(4)} & \colhead{(5)} & \colhead{(6)} & \colhead{(7)} 
}
\startdata
\\
grouping min 5 & $1.7 \pm 0.6$ & $0.8 \pm 0.1$ & $2.2 \pm 0.4$ & $1.0 \pm 0.3$ & $15.1 \pm 7.7$ & $ 49.71/68$\\
grouping min 15 & $1.6 \pm 0.6$ & $ 0.9 \pm 0.1$ & $2.2 \pm 0.4$ & $1.0 \pm 0.3$ & $15.2 \pm 7.8$ & $ 18.46/23$\\
\\
\enddata
\tablecomments{Col(1) --- model fit; Col(2) --- intrinsic hydrogen column density; Col(3--4) --- temperature and normalization of the collisionally ionized thermal plasma; Col(5--6) --- photon index and normalization of the power-law emission component; Col(7) --- fitting statistics.\\
All the models include Galactic hydrogen column density $N_{\rm H,\,Gal} = 1.83 \times 10^{20}$\,cm$^{-2}$; thermal component assumes solar abundance.\\
$^{\dagger}$ The \texttt{XSPEC} normalizations are: (a) for the APEC component $10^{-14} (1+z)^2 n_e n_{\rm H} V/4\pi d_{\rm L}^2$, assuming uniform ionized plasma with electron and H number densities $n_e$ and $n_{\rm H}$, respectively, and volume $V$, all in cgs units; (b) for the power-law component photons/keV/cm$^2$/s at 1\,keV.}
\end{deluxetable}

\end{document}